\documentclass[preprint,aps,physrev]{revtex4-2}
\usepackage[utf8]{inputenc}
\usepackage[british]{babel}

\usepackage{natbib}
\usepackage{graphicx}
\usepackage{subeqnarray}
\usepackage{bm}
\usepackage{amsmath, amssymb}
\usepackage{lineno}
\usepackage{amsfonts}
\usepackage{hyperref}
\usepackage{marginnote}

\providecommand\be{\begin{equation}}
\providecommand\ee{\end{equation}}

\providecommand\ba{\begin{align}}
\providecommand\ea{\end{align}}

\providecommand\bsub{\begin{subequations}}
\providecommand\esub{\end{subequations}}

\newcommand{\ve}[1]{\mbox{\boldmath $#1$}}
\newcommand{\bn}{\bm{\nabla}}

\newcommand{\bu}{\bm{u}}
\newcommand{\bx}{\bm{x}}

\begin{document}
\title{Fluid mechanics of free subduction on a sphere, 1: The axisymmetric case}
\date{\today}
\author{Alexander Chamolly}
\affiliation{Laboratoire de Physique de l’Ecole normale supérieure, ENS, Université PSL, CNRS, Sorbonne Université, Université de Paris, F-75005 Paris, France}
\affiliation{Institut Pasteur, Universit\'e de Paris, CNRS UMR3738, Developmental and Stem-Cell Biology Department, F-75015, Paris, France}
\author{Neil M. Ribe}
\email{ribe@fast.u-psud.fr}
\affiliation{Lab FAST, Universit\'e Paris-Saclay, CNRS, F-91405 Orsay, France}

\begin{abstract}
	
	To understand how spherical geometry influences the dynamics of gravity-driven subduction of oceanic lithosphere on Earth, we study a simple model of a thin and dense axisymmetric shell of thickness $h$ and viscosity $\eta_1$ sinking in a spherical body of fluid with radius $R_0$ and a lower viscosity $\eta_0$. Using scaling analysis based on thin viscous shell theory, we identify a fundamental length scale, the `bending length' $l_b$, and two key dimensionless parameters that control 
	the dynamics: the `flexural stiffness' $St = (\eta_1/\eta_0)(h/l_b)^3$ and the `sphericity number'
	$\Sigma= (l_b/R_0)\cot\theta_t$, where $\theta_t$ is the angular radius of the subduction trench.  To validate the scaling analysis, we obtain a suite of instantaneous numerical solutions using
	a boundary-element method based on new analytical point-force Green functions that satisfy free-slip boundary conditions on the sphere's surface. To isolate the effect of sphericity, we calculate the radial 
	sinking speed $V$ and the hoop stress resultant $T_2$ at the leading
	end of the subducted part of the shell, both 
	normalised by their `flat-Earth' values (i.e., for $\Sigma = 0$). 
	For reasonable terrestrial values of $\eta_1/\eta_0$ ($\approx$ several hundred), sphericity has a modest effect on $V$, which is reduced 
	by $< 7\%$
	for large plates such as the Pacific plate and by up to 34\% for
	smaller plates such as the Cocos and Philippine Sea plates. 
	However, sphericity has a much greater effect on $T_2$,
	increasing it by up to 64\% for large plates and 240\% for
	small plates. This result has important implications for the growth of
	longitudinal buckling instabilities in subducting spherical shells. 
	
\end{abstract}
\maketitle

\section{Introduction}

Subduction of oceanic lithosphere is a major component of Earth's plate tectonic cycle: it is the main source of the buoyancy that drives mantle convection; it is the principal process responsible for recycling oceanic crust and volatile species like water back into the mantle; it is the main driver of long-term continental deformation; and it generates most of the great earthquakes and explosive volcanoes on Earth. Subduction occurs because
oceanic lithosphere becomes denser as it cools moving away from the mid-ocean ridge where it
formed. The lithosphere is therefore gravitationally unstable, and sinks into the mantle via a 
subcritical instability in which its negative buoyancy is sufficient to overcome its internal resistance to bending \citep{mckenzie77agu}. 

Since the classic `corner flow' subduction flow model of \citet{mckenzie69},
hundreds of geodynamical models of subduction have been published,
too numerous to cite here. The majority of these models have used two- or three-dimensional
Cartesian geometry in which undeformed plates are flat. However, oceanic plates
are doubly-curved spherical shells, a fact that can be expected 
strongly to influence their mechanical behaviour. Whereas a flat plate supports a normal 
load by bending stresses alone, a doubly-curved shell can do so
by a combination of bending stresses and in-plane `membrane' stresses \citep{audoly10book}. 
Moreover, curvature stiffens a shell and renders it more resistant to bending than a flat plate. This is because in order to to change the intrinsic curvature of a surface, additional energy needs to be expended on stretching. Familiar examples are a magazine or newspaper that one rolls up to swat a fly, or an orange peel that resists being flattened.

That terrestrial subduction occurs on a sphere is of course no secret in the geodynamics 
community, and the effect of sphericity has been the subject of numerous studies. Early
models were purely geometrical, like the suggestion of \citet{frank68} that the shape of 
the lithosphere in subduction zones resembles that of a circular dent surrounding a hole in a 
deformed ping-pong ball. \citet{scholz70} and \citet{bayly82} suggested that a subducting 
spherical shell should buckle along the strike of the trench to accommodate the reduction
of the space available due to the spherical geometry as the shell penetrates deeper. 
\citet{laravie75} proposed a geometric model that predicts different degrees of lateral 
strain in the subducted part of the shell depending on its dip and its radius of curvature
at the surface. 
\citet{schettino12} reviewed the inadequacies of the ping-pong-ball model 
and proposed an alternative kinematic model. 

On the dynamical side, 
\citet{tanimoto97, tanimoto98} solved the equations for a 
normally loaded spherical elastic shell with negative buoyancy proportional to the (small)
normal displacement,
and concluded that the state of stress is strongly influenced by the spherical geometry. \citet{mahadevan10} 
used scaling analysis and numerical solutions for small-amplitude deformation of shallow spherical caps
to investigate the causes of the curvature and segmentation of subduction zones. For the case of 
an elastic shell on a thicker elastic foundation, they found a scaling law for the wavelength
of the edge instability (`dimpling') that occurs in response to a distributed radial body force. 
Finally, G.\ Morra and co-workers used the boundary-element method (BEM) to study 
large-amplitude subduction of viscous spherical shells, focusing
on the curvature of island arcs \citep{morra06geology}, subduction of single plates in a mantle with depth-dependent
viscosity, and interaction of multiple plates \citep{morra12}. 

While the models discussed above focus narrowly on the subduction process itself,
another approach is possible, namely time-dependent spherical thermal convection models in which
subduction is but one aspect of the numerically predicted global circulation. 
In most such models, subduction occurs in an unrealistic two-sided way,
with two adjoining plates subducting together \citep{coltice19sciadv}. However, 
\citet{schmeling08pepi} and 
\citet{crameri12grl} showed that realistic one-sided 
subduction could be obtained by adding to the top of the model domain 
a layer of low-viscosity fluid (`sticky air') to mimic a true free surface. From our
point of view, the disadvantages of global circulation models are their
high computational cost and relatively low spatial resolution, which typically
prevent scaling laws from being sought.

A crucial question for our modelling concerns the effective long-term rheology of 
oceanic lithosphere, which is the key factor controlling its resistance to 
deformation. On short time scales typical of earthquakes and of the Earth's
free oscillations, the lithosphere, like the rest of the solid Earth, behaves as
an elastic medium that transmits shear and compressional waves. 
However, on long time scales characteristic of mantle convection (tens of Ma), the
rheology of oceanic lithosphere is much more complicated \citep{karato08book}. 
Experimental rock mechanics shows that the rheology of the lithosphere under
conditions of strong bending is organised in three layers: brittle at shallow depths,
elastic at depths around the neutral surface where the fibre stress vanishes, 
and ductile at deeper depths 
due to the higher temperatures there. Moreover, ductile behaviour
can occur by three distinct mechanisms depending on temperature,
pressure, stress and grain size: diffusion creep with a Newtonian
rheology, dislocation creep with a generalised Newtonian (power-law) rheology,
and low-temperature Peierls (exponential) creep. There is clearly much to 
be learned by including all these deformation mechanisms in a single
realistic model (e.g.\ \citet{bessat20gji}). However, doing so tends to 
obscure the physical mechanisms at play and to prevent one from 
understanding scaling behaviour that a simpler model would reveal. 
Guided by these considerations, we have chosen to represent the rheology
of the lithosphere by a constant viscosity that is much larger than that
of the ambient mantle. In our view, this choice preserves the virtues of 
simplicity while embodying to lowest order the fact that the lithosphere
is rheologically stiff relative to its surroundings. 

In this paper, we present a simple model for the free (buoyancy-driven)
subduction of an axisymmetric viscous spherical shell in an ambient fluid with a lower viscosity. 
Because the model involves two fluids separated by a sharp interface, it is 
amenable to solution by the boundary-element method (BEM). An important advantage of the BEM is that 
numerical solutions are highly accurate and quick to obtain, making possible the
determination of clean quantitative scaling laws.  Our model and results are novel in several 
ways. First, we focus on large-amplitude subduction, and take into account the
self-consistent mechanical interaction between the highly deformed shell and the
less viscous ambient mantle. Second, we identify a fundamental length scale, the `bending
length', that characterises the flexural response of the loaded shell. And third, we
identify two key dimensionless parameters that govern the dynamics: a `flexural
stiffness' that determines which viscosity (shell or ambient) controls the sinking
rate of the shell, and a `sphericity number' that measures the importance of spherical geometry.

The paper is organised as follows. In \S~\ref{sec:model}, we present the model
problem and its geometrical and physical parameters. In \S~\ref{sec:scaling}
we perform a dimensional analysis and a physical scaling analysis to reveal the length scale and dimensionless parameters that govern the dynamics.
\S~\ref{sec:bem} describes our implementation of the boundary-element method. 
\S~\ref{sec:instantaneous} presents a suite of instantaneous BEM solutions, 
focusing on the effect of sphericity on the sinking speed of the shell and on
the longitudinal normal stress (hoop stress) within it.
\S~\ref{sec:timedep} presents illustrative time-dependent BEM solutions.
Finally, \S~\ref{sec:geophysical} estimates the magnitude of the sphericity
effect for six subduction zones in the Pacific Ocean basin. 

\section{Model}
\label{sec:model}

\begin{figure}
	\vspace*{-6cm}
	\begin{center}
		\includegraphics[width=0.8\linewidth]{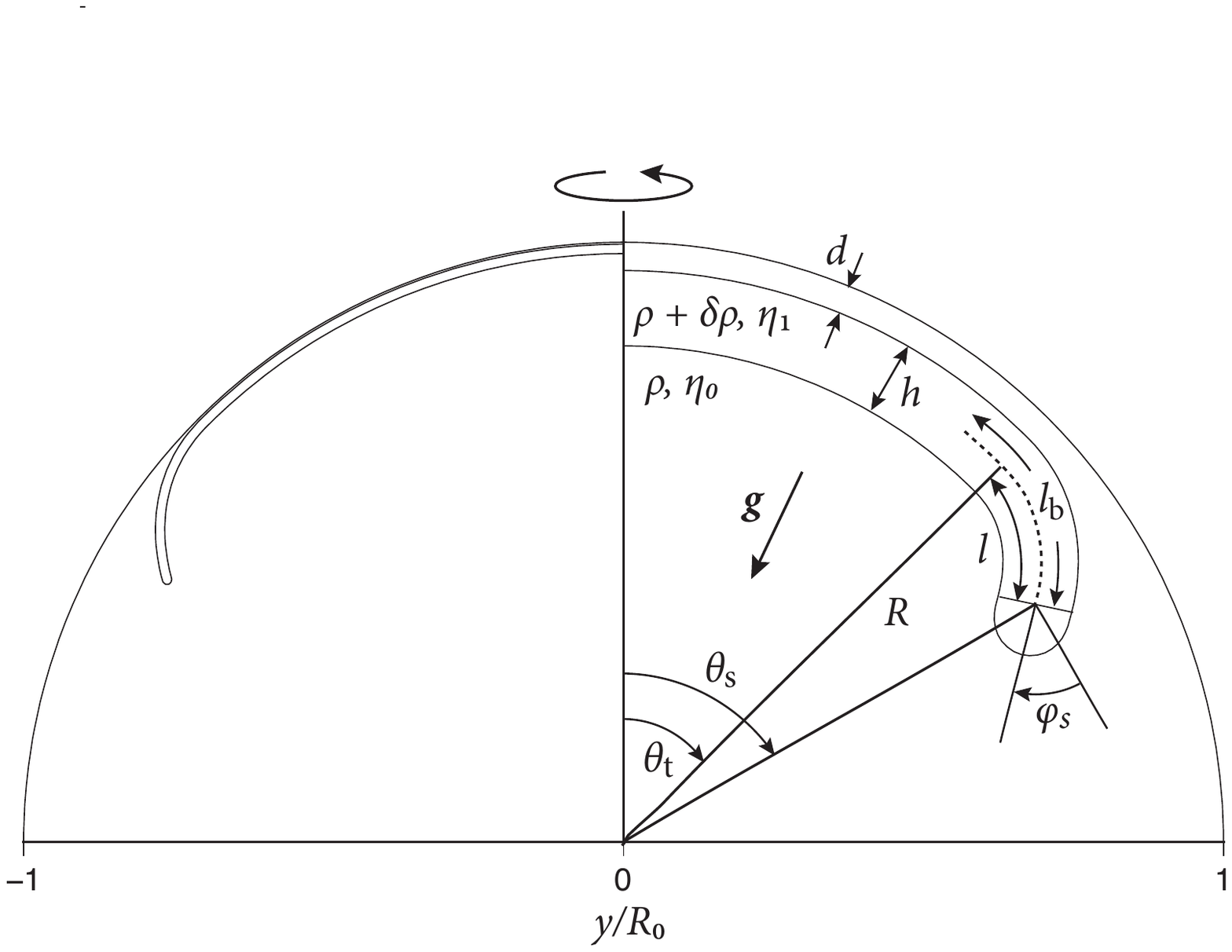}
	\end{center}
	\vspace*{-6cm}
	\caption{
		\label{fig_defsketch}
		Definition sketch of the model. Right portion (not to scale): an axisymmetric viscous shell with thickness
		$h$, density $\rho + \delta\rho$ and viscosity $\eta_1$ is immersed in a spherical 
		viscous planet with radius $R_0$, density $\rho$ and viscosity $\eta_0$. The
		outer surface of the planet is impermeable and free of shear traction. The shell comprises
		a spherical `plate' $\theta \leq \theta_t$ and a downward-dipping `slab' $\theta_t\leq \theta\leq\theta_s$.
		The lubrication layer above the plate has thickness $d$, and the radius of the unsubducted
		part of the plate's midsurface 
		(dotted line) is $R\equiv R_0 - h/2 - d$. The gravitational vector
		$\bm{g}$ is directed radially toward the centre of the planet. Left portion: the same shell
		drawn to a scale appropriate for the Earth, with  $h = 100$ km, $d=20$ km, and $R_0 = 6370$ km.
	}
\end{figure}

The model envisions an axisymmetric viscous shell with density 
$\rho_1=\rho+\delta\rho$ and viscosity $\eta_1$ immersed in a spherical
viscous planet with density $\rho_0=\rho$ and viscosity $\eta_0$
(figure~\ref{fig_defsketch}). The flow in both fluids is governed by the incompressible Stokes equations,
\begin{align}
	-\bm{\nabla}p+\eta_i\nabla^2\bm{u}+\rho_i\bm{g}=\bm{0},\quad\bm{\nabla}\cdot\bm{u}=0,
\end{align}
where $p$ is the pressure, $\bm{u}$ the velocity and $\rho_i\bm{g}$ is the gravitational force per unit volume.
The gravitational acceleration is directed radially toward the centre of the planet,
and its magnitude $g$ is treated as a constant for simplicity
($g$ is in any case nearly constant throughout the Earth's mantle). The outer surface $r = R_0$ of the planet is free-slip, i.e.\ the 
normal component of velocity and the shear traction both vanish there.
Our model lacks the
effectively inviscid core of radius $\approx 0.54 R_0$ that exists in the Earth. 
This neglect is deliberate: it gets rid of a non-essential dimensionless parameter
(the dimensionless core radius) 
that would complicate physical interpretation if it were retained. 
Neglecting the 
core also makes it possible to derive analytical Green functions for the flow
due to a point force, which in turn permit us to employ the
BEM to solve our problem.
We also neglect the planet's deviation from perfect sphericity due to the centrifugal forces 
associated with its rotation, which is typically small compared to $R_0$.

The first step in the BEM approach is to specify the shape of the shell. 
As indicated in figure~\ref{fig_defsketch},
the model shell comprises two pieces that join together
smoothly at a colatitude $\theta = \theta_t$. The inner piece $\theta\leq \theta_t$ is a circular spherical
cap of constant thickness $h$, separated from the free-slip
surface by a thin `lubrication layer' of thickness $d$. The radius of the midsurface
of the cap is therefore $R = R_0 - d - h/2$. The 
outer piece $\theta_t\leq\theta\leq\theta_s$ is a  
downward-dipping ring of arcwise length $l$ whose midsurface (dotted line in 
figure~\ref{fig_defsketch}) is located at the radius
\be
r(\theta) = R (1 - b \zeta^3 - c\zeta^4),
\quad
\zeta = \frac{\theta - \theta_t}{\theta_s - \theta_t}
\label{slabshape}
\ee
where $b$ and $c$ are constants. The thickness of the ring is
$h$ everywhere outside the rounded rim just beyond $\theta = \theta_s$.
Following
geophysical parlance, we shall call the cap $\theta < \theta_t$ the `plate',
the circle $\theta = \theta_t$ 
the `trench' and the downward-dipping ring $\theta_t\leq \theta \leq\theta_s$ the `slab' (these terms explain the choice of subscripts on $\theta_t$
and $\theta_s$).

The mathematical form of Eq.~\eqref{slabshape} ensures that the local slope and local
curvature (in the $\theta$-direction) of the slab match those of the plate at the trench $\zeta = 0$. 
The values of $b$ and $c$ are determined by imposing two additional constraints
on Eq.~\eqref{slabshape}. The first is that the dip $\varphi$ of the leading end of the slab's
midsurface
relative to the local horizontal is a specified value $\varphi_s$. The second 
constraint is that the curvature at the leading end of the 
midsurface is $-1/R$. The reason for imposing this constraint is that the 
slab's leading edge is free, i.e.\ the bending moment is zero there. Because
the bending moment in a viscous shell is proportional to the rate of change of curvature
of the midsurface,
the curvature at the leading edge of the slab should not deviate from the
value $-1/R$ for an undeformed spherical shell. Explicitly, the foregoing 
two constraints are
\be
\left. \frac{r}{\left(r^2 + (\partial_{\theta} r\right)^2)^{1/2}}\right |_{\zeta = 1} = \cos\varphi_s,
\label{slopeconstraint}
\ee
\be
\left. -\frac{r^2 + 2 (\partial_{\theta} r)^2 - r\partial^2_{\theta\theta} r}
{\left(r^2 + (\partial_{\theta} r\right)^2)^{3/2}}\right |_{\zeta = 1} = -\frac{1}{R}.
\label{curvconstraint}
\ee
Eqs.~\eqref{slopeconstraint} and \eqref{curvconstraint} are a rather
complicated quartic system of algebraic equations for $b$ and $c$, which
we solved numerically for given values of $R$, $\varphi_s$ and $\theta_s - \theta_t$.

A few words of explanation are in order concerning the lubrication layer above
the plate in figure~\ref{fig_defsketch}, which corresponds to the `sticky air' layer 
mentioned earlier. 
The function of this layer is twofold: it creates large normal stresses
that support the plate and prevent it from sinking; and it applies a negligible
shear stress to the upper surface of the plate. The latter surface
is thus effectively a free surface whose radial velocity is not constrained,
allowing the plate to bend in a realistic way. 
The advantage of the lubrication layer formulation is that it obviates the need to 
deal with the numerically troublesome
complexities of a true free surface, including the triple point at the trench where
the shell, the ambient fluid and the overlying ocean would meet.
\citet{li12} have shown that the predictions of Cartesian three-dimensional subduction models
with a lubrication layer agree quantitatively (not just qualitatively) with laboratory experiments in which the plate
is prevented from sinking by surface tension. This shows that two different mechanisms for 
preventing the sinking of the plate lead to identical predictions, inspiring confidence 
in the lubrication layer approach.

A final length scale that we shall need is the `bending length' $l_b$, which plays a 
crucial role in the subsequent scaling analysis. 
When the Stokes equations are solved for an immersed shell like
that shown in figure~\ref{fig_defsketch}, the resulting velocity field will be
non-zero everywhere inside the shell. Because the shell is thin, however,
that velocity field can be described as a combination of bending and stretching.
Based on previous experience with
two-dimensional shells \citep{ribe10}, we anticipate that the rate of deformation
will be dominated by bending throughout the slab and in an adjoining portion
of the plate. We call the arcwise length of this part
of the shell the `bending length' $l_b$. In geophysical terms,
the bending length is the sum of the slab length $l$ and the length $l_b - l$
of the near-trench portion of the plate where significant upward
motion (`flexural bulging') occurs. The bending length is indicated
schematically in the right-hand portion of figure~\ref{fig_defsketch}. 
Because the bending length is a dynamic length scale that arises physically from
solving the Stokes equations, it is of a fundamentally
different character than the geometric length scales $h$, $d$ and $l$. 

\section{Dimensional and scaling analysis}
\label{sec:scaling}

\subsection{Dimensionless groups}

We begin by performing a dimensional analysis to get an idea of the `size' of the problem we face.
The first task is to choose a characteristic output parameter of the model to serve as the target of the 
analysis. Following our earlier studies of subduction in two- and three-dimensional Cartesian geometry
\citep{ribe10, li12}, we choose the vertical (i.e.\ radial) component $V$ of the velocity of the
leading edge of the slab. This choice is based on the fact that in Cartesian geometry, $V$ satisfies a simpler
scaling law than other possible choices such as the norm or the horizontal component of the 
leading-edge velocity \citep{ribe10}. 

Referring to figure \ref{fig_defsketch}, we see that $V$ depends on the planet radius $R_0$, the shell
thickness $h$, the lubrication layer thickness $d$, the slab length $l$, the slab dip $\varphi_s$,
the angle $\theta_t$ subtended by the plate,  the mantle viscosity $\eta_0$, the shell viscosity
$\eta_1$, and the buoyancy $g\delta\rho$. The gravitational acceleration and the densities of 
the shell and the mantle appear only in the combination $g\delta\rho$ because all velocities
in gravity-driven Stokes flow are linearly proportional to the driving buoyancy force. 

The foregoing list
comprises ten parameters, three of which have independent dimensions. Buckingham's $\Pi$-theorem \citep{buckingham14}
therefore tells us that seven independent dimensionless groups can be formed from our original
ten parameters. Incorporating $V$ into only one of the groups for convenience, 
we see that the scaling law satisfied by $V$ must have the general form
\be
\frac{V \eta_0}{hl g\delta\rho} = f_0
\left(
\frac{h}{R_0},
\frac{h}{l},
\frac{d}{h},
\frac{\eta_1}{\eta_0},
\theta_t,
\varphi_s
\right)
\label{scalelawgeneral}
\ee
where $f_0$ is an unknown function. The seven dimensionless groups could of course have been
defined differently, but the choices in Eq.~\eqref{scalelawgeneral} are as good as any. 

While dimensional analysis has reduced the number of independent parameters substantially,
we are still faced with an unknown function of six arguments. Characterising such a function
completely using numerical simulations is not practical, and would not provide any
insight even if it were. However, we have not yet 
exploited the physics of our problem, which we now do using a scaling
analysis based on the theory of thin viscous shells. 

\subsection{Physical scalings}
\label{sec:physscaling}

We now turn to the scaling analysis, focusing on the three forces that 
act on the bending portion (of length $l_b$) of the shell. These are the buoyancy force $F_b$, the external viscous force $F_{ext}$ exerted on the shell by the surrounding fluid, and the internal viscous force $F_{int}$ that resists bending of the shell.  The goal is first to find scalings for each of these in terms of the problem parameters, and then use them to find the scaling of the sinking rate $V$ of the slab.

The buoyancy force acting on the slab (per unit circumferential length) scales as
\be
F_b\sim l h g\delta\rho.
\label{fbscale}
\ee
The buoyancy force is proportional to $l$ rather than $l_b$ because
the buoyancy of the shell in the flexural bulging region is exactly 
compensated by normal stresses in the overlying lubrication layer \citep{ribe10}.

The viscous traction (force per unit area) applied to the shell
by the outer fluid is $\sigma_{ext} \sim \eta_0 V/l_b$, where
$V > 0$ is the downward radial velocity at the tip of the slab. Integrating
this over the bending length, we find the external force per unit
circumferential length acting on the shell:
\be
\label{fextscale}
F_{ext}\sim \eta_0 V.
\ee

Now we turn to the internal viscous force $F_{int}$ that resists deformation
of the shell. This force is just the integral $N_1$ of the radial
shear stress acting on a cross-section of the shell. To estimate this
force, we exploit the theory of thin elastic shells \citep{novozhilov59}
together with the Stokes-Rayleigh analogy between incompressible
elasticity and slow viscous flow \citep{rayleigh45}. According to
this analogy, elastic shell theory can be transformed into its 
viscous equivalent by interpreting displacements as velocities
and making the transformations $E\to 3\eta$ and $\sigma\to 1/2$,
where $E$ is Young's modulus, $\eta$ is the viscosity
and $\sigma$ is Poisson's ratio.
From Eq.~(7.8) of
\citet{novozhilov59}, we have
\be
F_{int} \equiv N_1 = \frac{1}{A_1 A_2} 
\left[
\partial_{\theta} (A_2 M_1) - M_2\partial_{\theta} A_2
\right],
\label{n1gen}
\ee
where 
\be
A_1 = \left[ r^2 + (\partial_{\theta} r)^2\right]^{1/2}
\quad
\mathrm{and}
\quad
A_2 = r\sin\theta
\label{lamepars}
\ee
are the Lam\'e parameters of the deformed axisymmetric shell and
$M_1$ and $M_2$ are bending moments.  The explicit expressions for these are
\be
M_1 = \frac{\eta h^3}{3}\left( \kappa_1 + \frac{1}{2}\kappa_2 \right),
\quad
M_2 = \frac{\eta h^3}{3}\left( \kappa_2 + \frac{1}{2}\kappa_1 \right)
\label{m1m2def}
\ee
where 
\be
\kappa_1 = - \frac{1}{A_1}\partial_{\theta} 
\left(
\frac{1}{A_1}\partial_{\theta} W + K_1 U
\right),
\quad
\kappa_2 = - \frac{1}{A_1 A_2}\partial_{\theta} A_2
\left(
\frac{1}{A_1}\partial_{\theta} W + K_1 U
\right)
\label{roccurv}
\ee
and $U$ and $W$ are the velocity components tangential to and normal
to the midsurface, respectively. The quantity $K_1$ in Eq.~\eqref{roccurv} is the curvature
of a line of constant longitude; its explicit expression is the 
left-hand side of Eq.~\eqref{curvconstraint}. Physically, the 
quantities $-\kappa_1$ and $-\kappa_2$ are the rates of change of 
curvature of the midsurface in the $\theta$- and $\phi$-directions,
respectively, due to deformation by pure bending without stretching
(see \citet{novozhilov59} pp. 25-26 for a discussion). 

For scaling purposes, we approximate $A_1$
and $A_2$ as their values for an undeformed spherical surface of
radius $R$, where (to recall) $R$ is the radius of the
midsurface of the plate. Thus we have $A_1\approx R$ and $A_2\approx R\sin\theta$. 
Moreover, we neglect the terms $K_1 U$ in Eq.~\eqref{roccurv}, which
our subsequent numerical solutions show to be small compared to ${A_1}^{-1}\partial_{\theta} W$. 
We then obtain
\be
\kappa_1\approx - \frac{1}{R^2}\partial^2_{\theta\theta} W,
\quad
\kappa_2\approx - \frac{\cot\theta}{R^2}\partial_{\theta} W.
\ee
Furthermore,  Eq.~\eqref{n1gen} takes the form 
\be
F_{int} = \frac{1}{R} (M_1 - M_2)\cot\theta + \frac{1}{R}\partial_{\theta} M_1.
\label{n1simp}
\ee
Now using the scaling $(1/R) \partial_{\theta}\sim1/l_b$, we find
\be
M_1, M_2, M_1 - M_2\sim \frac{\eta_1 h^3 V}{l_b^2} \left\langle1, \frac{l_b \cot\theta}{R}\right\rangle,
\label{m1m2scale}
\ee
where the quantities inside $\langle\rangle$ indicate two terms with different scalings. Ignoring the small difference between $R$ and $R_0$ and 
choosing $\theta_t$ as a representative value of $\theta$, we define a dimensionless
`sphericity number'
\be
\Sigma = \frac{l_b}{R_0}\cot\theta_t.
\ee
Eqs.~\eqref{n1simp} and \eqref{m1m2scale} now imply 
\be
F_{int} \sim \eta_0 V f_1(St ,\Sigma)
\label{fintscale}
\ee
where
\be
St = \frac{\eta_1}{\eta_0} \left( \frac{h}{l_b}\right)^3
\label{stdef}
\ee
is a dimensionless `flexural stiffness'. Eq.~\eqref{stdef} is
identical to the flexural stiffness for subduction of initially
flat plates in
two- and three-dimensional Cartesian geometry \citep{ribe10, li12}. 

Having obtained the scalings for the three forces in Eqs.~\eqref{fbscale}, \eqref{fextscale} 
and \eqref{fintscale}, we proceed to determine the scaling of the sinking speed $V$.
Balancing $F_b$ and $F_{ext}$ yields
the Stokes velocity scale 
\be
V_{Stokes} = \frac{l h g\delta\rho}{\eta_0}.
\label{vstokesdef}
\ee
We expect the normalised sinking speed $V/V_{Stokes}$ to be a function of the 
ratio $F_{int}/F_{ext}$ and of the two angles $\theta_t$ (plate radius) and $\varphi_s$ (slab dip)
whose purely geometrical influence cannot be captured by scaling analysis.  
We therefore obtain a scaling law of the general form
\be
\frac{V}{V_{Stokes}}= f_2(St, \Sigma, \theta_t, \varphi_s),
\equiv f_2\left( 
\frac{\eta_1}{\eta_0} \frac{h^3}{l_b^3},
\frac{l_b}{R_0}\cot\theta_t,
\theta_t, \varphi_s \right)
\label{vvstokesscale}
\ee
where $f_2$ is an unknown function. The ratio $d/h$ does not appear on the 
right-hand side of Eq.~\eqref{vvstokesscale} because its only effect is to modify the bending
length slightly \citep{ribe10}. By introducing the intermediate variable
$l_b$, our scaling analysis has succeeded in reducing the function Eq.~\eqref{scalelawgeneral}
of six dimensionless arguments to Eq.~\eqref{vvstokesscale},
which involves only four arguments. 

In the sequel we shall have occasion to compare our numerical
predictions with those for a reference `flat-Earth' limit in
which spherical effects are absent. This limit corresponds
to $\Sigma\to 0$, and can be achieved in two ways. The first is
to note that $\Sigma$ may be written as
\be
\Sigma = \frac{l_b}{L}\epsilon\cot\epsilon
\quad
\mathrm{where}
\quad
\epsilon = \frac{L}{R_0}
\ee
and $L$ is the arcwise radius of the trench.
Because $\lim_{\epsilon\to 0}\epsilon\cot\epsilon = 1$,
we obtain $\Sigma\to 0$ in the double limit $L/R_0\to 0$,
$l_b/L\to 0$. However, this limit is not easily accessible
numerically, and so we define the flat-Earth limit in a
different way as $\theta_t = \pi/2$. This definition may not be immediately obvious,
because it corresponds to a plate in the form of a complete hemisphere. However, the analogy becomes clearer when we consider that for a
hemispherical plate the principal normal vector to the trench is perpendicular to the shell. The two-dimensional Cartesian (``true flat-Earth") case \citep{ribe10} is then recovered smoothly in the limit where the shell is vanishingly thin compared to the radius of curvature of the edge, which is approximately the radius of the Earth. The situation is fundamentally different for a spherical cap with $\theta_t < \pi/2$, for which the principal normal has a component parallel to the shell's midsurface. We show in \S\ref{sec:bbl} and figure \ref{fig_vvstokes_vs_st0} that the sinking speed $V(St)$ in the hemispherical limit $\theta_t = \pi/2$ agrees closely with the two-dimensional Cartesian prediction for values of the 
flexural stiffness $St$ relevant to the Earth. We may therefore
use $\theta_t=\pi/2$ as a physically sound proxy for the flat-Earth limit. The function
Eq.~\eqref{vvstokesscale} then simplifies to 
\be
\left(
\frac{V}{V_{Stokes}}
\right)_{flat} = f_2\left (St, 0, \frac{\pi}{2}, \varphi_s\right )\equiv f_3(St,\varphi_s).
\label{vvstokesflat}
\ee


\section{Boundary-element method}
\label{sec:bem}

Our starting point is the general boundary-integral representation for the flow
generated by a buoyant drop with density excess $\delta\rho$
and viscosity $\eta_1$ immersed in another fluid with viscosity $\eta_0$. Let $V_1$
denote the drop, $V_0$ denote the external fluid, and $S$ denote the interface 
between them. Moreover, let $u_j^{(m)}(\bm{x})$ be the velocity in
volume $V_m$. Then the flow inside and outside the drop is governed by
the integral equation \citep{pozrikidis90, manga93jfm}
\begin{align}\label{inteqn1}
	\chi_0(\bm{x}_0) u_j^{(0)}(\bm{x}_0)  + \gamma \chi_1(\bm{x}_0) u_j^{(1)}(\bm{x}_0)
	=& -\frac{\delta\rho}{\eta_0}\int_S (\bm{g}(\bm{x})\cdot\bm{x})G_{ij}(\bm{x}, \bm{x}_0) n_i(\bm{x})\mathrm d S(\bm{x})\nonumber\\
	&+ (1-\gamma)\int_S u_i(\bm{x}) T_{ijk}(\bm{x}, \bm{x}_0) n_k(\bm{x})\mathrm d S(\bm{x}).
\end{align}
Here $G_{ij}$ and $T_{ijk}$ are Green functions for the velocity and stress, respectively, at the point
$\bm{x}_0$ generated by a point force acting at $\bm{x}$.
Also, $\chi_0(\bm{x}_0) = 1$ if $\bm{x}_0$ is in $V_0$, $1/2$ if $\bm{x}_0$ is right on $S$,
and 0 if $\bm{x}_0$ is in $V_1$. $\chi_1(\bm{x}_0)$ is defined similarly but with the subscripts
0 and 1 interchanged. The velocity $\bm{u} (\bm{x}_0)$ on $S$ satisfies an integral
equation obtained from Eq.~\eqref{inteqn1} by setting $\chi_0 = \chi_1 = 1/2$ and applying
the velocity matching condition $u_j^{(0)}(\bm{x}_0) = u_j^{(1)}(\bm{x}_0) = u_j(\bm{x}_0)$, yielding
\begin{align}\label{inteqn2}
	\frac{1}{2} (1 + \gamma) u_j(\bm{x}_0) = &-\frac{\delta\rho}{\eta_0}\int_S (\bm{g}\cdot\bm{x})G_{ij}(\bm{x}, \bm{x}_0) n_i(\bm{x})\mathrm d S(\bm{x})\nonumber\\
	&+ (1-\gamma)\int_S u_i(\bm{x}) T_{ijk}(\bm{x}, \bm{x}_0) n_k(\bm{x})\mathrm d S(\bm{x}).
\end{align}
Next we subtract the singularities of the two integrands in Eq.~\eqref{inteqn2} following the procedure outlined 
in \S6.4 of  \citet{pozrikidis92}, which yields
\bsub
\begin{align}
	\int_S (\bm{g}\cdot\bm{x})G_{ij}(\bm{x}, \bm{x}_0) n_i(\bm{x})\mathrm d S(\bm{x})
	= \int_S [\bm{g}(\bm{x})\cdot\bm{x} - \bm{g}(\bm{x}_0) \cdot \bm{x}_0 ]G_{ij}(\bm{x}, \bm{x}_0) n_i(\bm{x})\mathrm d S(\bm{x}),
\end{align}
and
\begin{align}\label{singsubtract}
	&\int_S u_i(\bm{x})T_{ijk}(\bm{x}, \bm{x}_0) n_k(\bm{x})\mathrm d S(\bm{x})\nonumber\\
	=& - \frac{1}{2} u_j (\bm{x}_0) + \int_S [u_i(\bm{x}) - u_i(\bm{x}_0)] T_{ijk}(\bm{x}, \bm{x}_0) n_k(\bm{x})\mathrm d S(\bm{x}).
\end{align}
\esub

Substituting Eq.~\eqref{singsubtract} into Eq.~\eqref{inteqn2} we obtain
\begin{align}\label{inteqn3}
	u_j(\bm{x}_0) =& - \frac{\delta\rho}{\eta_0}\int_S [\bm{g}(\bm{x})\cdot\bm{x} - \bm{g}(\bm{x}_0) \cdot \bm{x}_0 ] G_{ij}(\bm{x}, \bm{x}_0) n_i (\bm{x}) \mathrm d S(\bm{x})\nonumber\\
	&+ (1 -\gamma)\int_S [u_i(\bm{x}) - u_i(\bm{x}_0)] T_{ijk} (\bm{x}, \bm{x}_0) n_k(\bm{x})\mathrm d S(\bm{x}).
\end{align}

We now set $\bm{g} = - g\bm{e}_r$ and non-dimensionalise all lengths by $R_0$ and all velocities by 
$g\delta\rho R_0^2/\eta_0$ (we do not non-dimensionalise using the velocity scale $V_{Stokes}$
because it contains the variable slab length $l$). Eq.~\eqref{inteqn3} then takes the form
\begin{subequations}\label{inteqn4}
	\begin{align}
		u_j(\bm{x}_0) &= - \mathcal S_j(\bm{x}_0)+ (1- \gamma) \mathcal D_j(\bm{x}_0) ,\\
		\mathcal S_j(\bm{x}_0) &= - \int_S [\bm{e}_r(\bm{x}) \cdot \bm{x} - \bm{e}_r(\bm{x}_0) \cdot\bm{x}_0] n_i(\bm{x}) G_{ij}(\bm{x}, \bm{x}_0)\mathrm d S(\bm{x}),
		\label{singpot1}\\
		\mathcal D_j(\bm{x}_0) &= \int_S [u_i(\bm{x}) - u_i(\bm{x}_0)] T_{ijk} (\bm{x}, \bm{x}_0) n_k(\bm{x})\mathrm d S(\bm{x}),\label{doubpot1}
	\end{align}
\end{subequations}
where all variables are now dimensionless.
The quantity $\mathcal S_j(\bm{x})$ is called the single-layer potential, and $\mathcal D_j(\bm{x})$ is the double-layer potential.

The Green functions $G_{ij}$ and $T_{ijk}$ for flow inside a sphere
are most readily derived by splitting the point force into a radial component normal to the sphere's surface
and a transverse component tangential to it (Appendix \ref{app:green}) . It then proves convenient to work with
mixed-index Green functions $\hat G_{i\alpha}(\bm{x}, \bm{x}_0)$ and 
$\hat T_{i\alpha k}(\bm{x}, \bm{x}_0)$, where the index $\alpha$ can take on either
of the two values $r$ or $\theta$ according to whether the unit force points in the
radial or the transverse direction at the location $\bm{x}_0$. Quantities involving 
mixed indices are indicated by a superposed hat. 
The key advantage of replacing the Cartesian index $j$ by the spherical polar index $\alpha$ is that depending on the
latter's value we need use only the Green function for either a radial force or a transverse one to calculate $\hat{\bm{G}}$ and 
hence the corresponding component of the single- and double-layer potentials. 

Consider first the single-layer potential Eq.~\eqref{singpot1}, which we rewrite as
\begin{align}
	\hat{\mathcal S}_\alpha(\bx_0)=\int_D (r_0-r)\hat{G}_{i\alpha}(\bx,\bx_0)n_i(\bx)\, \text{d}S(\bx).
	\label{singpot2}
\end{align}
Next, we specialise Eq.~\eqref{singpot2} to axisymmetric flow by integrating over the azimuthal angle $\phi$.
Following the notation of \citet{pozrikidis92}, we introduce cylindrical polar coordinates $(x,\sigma,\phi)$
such that the cylindrical axis $\sigma = 0$ corresponds to the Cartesian $x$-axis.
We then set $\mathrm d S = \sigma \mathrm d\phi\mathrm d l$, where $\mathrm d l$ is the differential arclength
of the trace of the contour $C$ in any azimuthal plane. We also note that the Cartesian components of the 
normal vector $\bm{n}$ are $(n_x, n_{\sigma}\cos\phi, n_{\sigma}\sin\phi)$. To perform the integration,
we suppose for definiteness that the field point $\bm{x}_0$ lies in the $x$-$y$ plane where $\phi_0 = 0$. The
single-layer potential can then be written as
\begin{align}
	\hat{\mathcal S}_\alpha(\bx_0)=\int_C (r_0-r)\hat{\mathcal{M}}_{\alpha\beta}(\bx,\bx_0)n_\beta(\bx)\, \text{d}l(\bx),
	\label{sglintegral}
\end{align}
where
\begin{align}
	\begin{pmatrix}
		\hat{\mathcal{M}}_{rx} & \hat{\mathcal{M}}_{r\sigma} \\
		\hat{\mathcal{M}}_{\theta x} & \hat{\mathcal{M}}_{\theta\sigma}
	\end{pmatrix}=
	\sigma\int_0^{2\pi}\begin{pmatrix}
		\hat{G}_{xr} & \hat{G}_{yr}\cos\phi+\hat{G}_{zr}\sin\phi \\
		\hat{G}_{x\theta} & \hat{G}_{y\theta}\cos\phi+\hat{G}_{z\theta}\sin\phi
	\end{pmatrix}
	\,\text{d}\phi.
\end{align}
Note that only the radial Green function is necessary to calculate $\hat{\mathcal{M}}_{rx}$ and $\hat{\mathcal{M}}_{r\sigma}$, and hence the radial component $\hat{\mathcal S}_r$ of the single layer potential. The situation for the tangential component is similar.

The procedure for the double-layer potential is analogous. The result is
\begin{align}
	\hat{\mathcal D}_\alpha(\bx_0) =\int_C \left[ u_\beta(\bx)\hat{\mathcal{Q}}_{\alpha\beta \gamma}(\bx,\bx_0) - 
	u_\beta(\bx_0)\hat{\mathcal{P}}_{\alpha\beta \gamma}(\bx,\bx_0)\right] n_\gamma(\bx)\, \text{d}l(\bx),
	\label{dblintegral}
\end{align}
where the indices $\beta$ and $\gamma$ take cylindrical values $\{x,\sigma\}$ and
the tensors $\hat{\mathcal{P}}$ and $\hat{\mathcal{Q}}$ are defined as
\begin{subequations}	
	\begin{align}
		\begin{pmatrix}
			\hat{\mathcal{P}}_{\alpha xx} & \hat{\mathcal{P}}_{\alpha x\sigma} \\
			\hat{\mathcal{P}}_{\alpha \sigma x} & \hat{\mathcal{P}}_{\alpha \sigma\sigma}
		\end{pmatrix}&=
		\sigma\int_0^{2\pi}\begin{pmatrix}
			\hat{T}_{x\alpha x} & \hat{T}_{y\alpha x}\cos\phi+\hat{T}_{z\alpha x}\sin\phi \\
			\hat{T}_{x\alpha y} & \hat{T}_{y\alpha y}\cos\phi+\hat{T}_{z\alpha y}\sin\phi
		\end{pmatrix}
		\,\text{d}\phi,\\
		\hat{\mathcal{Q}}_{\alpha xx} &= \hat{\mathcal P}_{\alpha x x},
		\quad
		\hat{\mathcal{Q}}_{\alpha x\sigma}= \hat{\mathcal{Q}}_{\alpha \sigma x} = 
		\hat{\mathcal{P}}_{\alpha x\sigma},\\
		\hat{\mathcal{Q}}_{\alpha \sigma\sigma} &= \sigma\int_0^{2\pi} \left( \hat{T}_{y\alpha  y}\cos^2\phi
		+\hat{T}_{z\alpha  z}\sin^2\phi+\hat{T}_{y\alpha  z}\sin2\phi\right)\mathrm d\phi.
	\end{align}
\end{subequations}
Again, only the radial (tangential) Green function is needed to calculate the radial (tangential) 
component of the double-layer potential at $\bx_0$.

To evaluate the single- and double-layer integrals Eq.~\eqref{sglintegral} and Eq.~\eqref{dblintegral}, 
we discretise the contour $C$
using three-node curved elements $C_n$  $(n = 1, 2, ... , N)$, over each of which
$\ve x$ and $\ve u$ vary  as
\be
\ve x(\xi) = \sum_{m=1}^3\phi_m(\xi) \ve x_m,
\quad
\ve u(\xi) = \sum_{m=1}^3\phi_m(\xi) \ve u_m,
\label{quadvar}
\ee 
where $\ve x_m$ are the  (known) nodal coordinates,
$\ve u_m$ are the  (unknown) nodal velocities, and
$\phi_m(\xi)$ are quadratic basis functions defined on a 
master element $\xi\in [-1,1]$.  Substitution of Eq.~\eqref{quadvar}
into Eq.~\eqref{sglintegral} and Eq.~\eqref{dblintegral}
with $\ve x_0\in C$ transforms the integrals over $C$ into 
sums of integrals over the elements $C_n$, each of which
is evaluated on $\xi\in [-1,1]$ using 6-point Gauss-Legendre
quadrature. The resulting system of $4N + 2$ coupled linear equations
was solved 
using LU decomposition and back-substitution \citep{press92},
yielding the nodal velocities $\ve u_m$ with fourth-order accuracy.
The correctness of the instantaneous solution for the velocity was verified 
against an analytical solution for two concentric spheres with different
viscosities (Appendix \ref{app:concentric}). 
For the few time-dependent cases we considered, the positions of all material
points $\ve x\in C$ were advanced at each time step using explicit Euler stepping.

\section{Instantaneous solutions}
\label{sec:instantaneous}

\subsection{General features}

Because inertia is negligible in planetary mantles, the evolutionary history of a subducting shell is 
nothing more than a sequence of quasi-static configurations whose dynamics are determined entirely by 
the shell's instantaneous shape. It therefore makes sense first to study the quasi-static dynamics of a 
shell as a function of its viscosity ratio and shape, without the added complexity of the (purely kinematic) 
time evolution.
Thus instead  of  regarding the model parameters shown in figure~\ref{fig_defsketch} ($\theta_t$, $\theta_s$, $\varphi_s$ etc.) as  
initial values for a time-dependent simulation, we treat them as free geometrical parameters that can be varied to 
represent a wide range of different shell shapes at some arbitrary instant in time. 

\begin{figure}
	\includegraphics[width=0.8\linewidth]{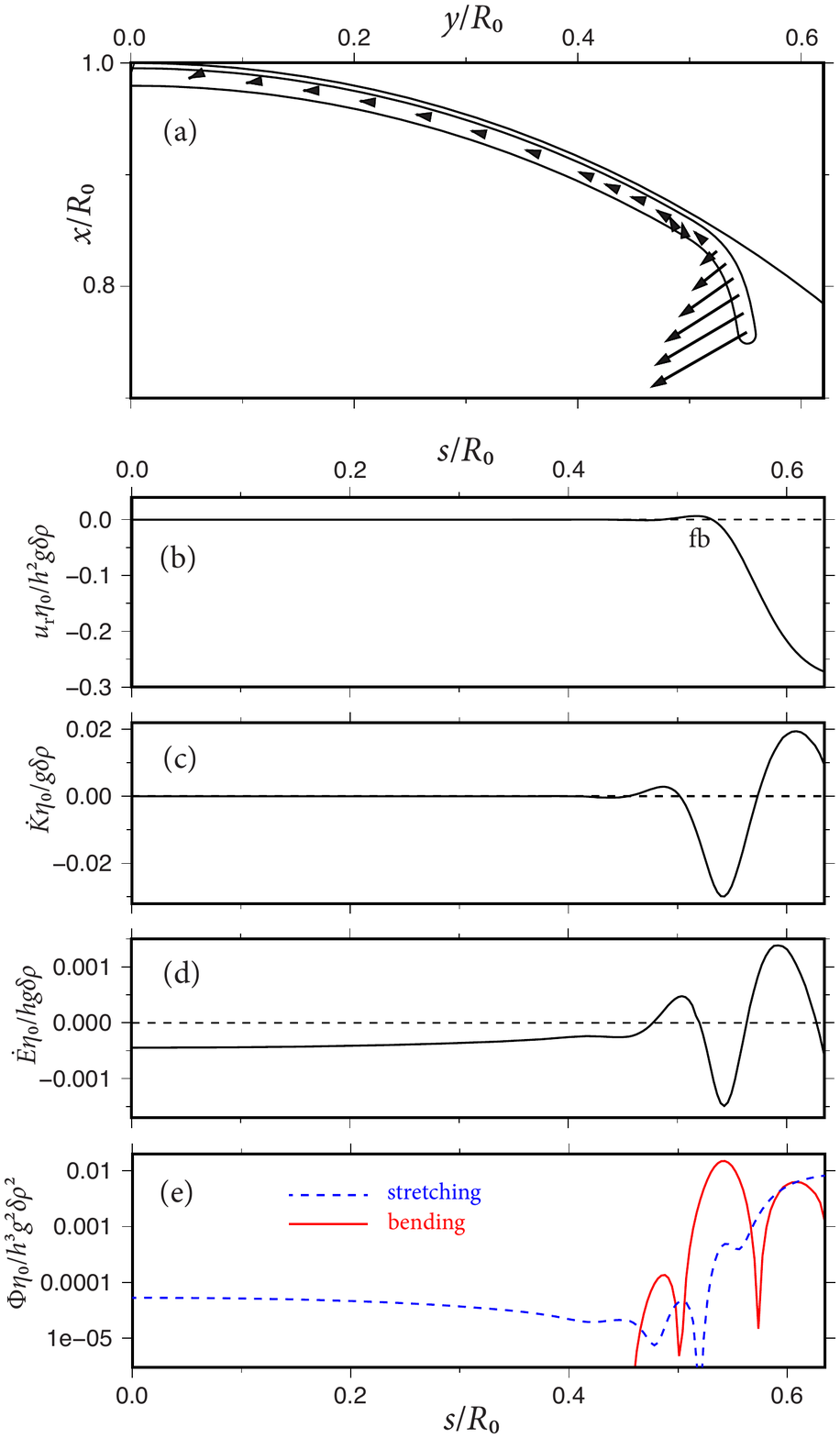}
	\vspace*{-3.0cm}
	\caption{
		\label{fig_std_gam1e2}
		Dynamics of an illustrative instantaneous subduction solution with $\theta_t = 30^{\circ}$, $\theta_s = 36^{\circ}$,
		$h/R_0 = 0.0157$, $d/h = 0.3$, $\varphi_s = 45^{\circ}$ and $\gamma = 100$. (a) Geometry of the shell with midsurface velocity vectors. 
		The longest arrow has magnitude $0.299 h^2 g\delta\rho/\eta_0$. (b) Radial velocity $u_r$. The label ``fb" indicates a region 
		of upward radial velocity corresponding to flexural bulging seaward of the trench. (c) Bending
		rate $\dot K$ and (d) stretching rate $\dot E$ as functions
		of arclength $s$ along the midsurface. (e) Rates of viscous dissipation of energy due to bending (red) and stretching (blue)
		as functions of $s$.
	}
\end{figure}

To begin, we examine the instantaneous flow of a reference shell with the parameters $\theta_t = 30^{\circ}$, 
$\theta_s = 36^{\circ}$, $\varphi_s = 45^{\circ}$, $h/R_0 = 0.0157$, $d/h = 0.3$, and $\gamma = 100$.
Figure~\ref{fig_std_gam1e2}a shows the velocity on the shell's midsurface, calculated by averaging
the velocities on the upper and lower surfaces of the shell adjoining each midsurface point. The 
slab moves downward and backward, while the nearly horizontal velocity within the plate is of 
much smaller magnitude. Figure~\ref{fig_std_gam1e2}b shows the radial velocity as a function of
arclength along the midsurface. The most interesting feature is an interval of upward radial 
velocity $s/R_0\in [0.49, 0.53]$. This corresponds to the so-called `flexural bulge' of the ocean
floor that is observed to exist seaward of many trenches on Earth. 

To proceed further with our analysis, we now exploit the fact that the shell is a thin object whose
rate of deformation can be described as a combination of bending and stretching. 
According to the thin-shell constitutive relation Eq.~\eqref{m1m2def},
the rate of 
longitudinal bending of the shell's midsurface is measured by the effective bending rate
\be
\dot K = - \left(\kappa_1 + \frac{1}{2}\kappa_2\right)
\ee
where $\kappa_1$ ad $\kappa_2$ are defined by Eq.~\eqref{roccurv}. Similarly, the effective 
rate of longitudinal stretching of the midsurface is
\be
\dot E = \epsilon_1 + \frac{1}{2}\epsilon_2
\ee
where
\be
\epsilon_1 =  \frac{1}{A_1}\partial_{\theta} U - K_1 W,\quad
\epsilon_2 = \frac{1}{A_1 A_2} U\partial_{\theta} A_2 - K_2 W
\label{eps12}
\ee
and $K_2$ is the second principal curvature of the midsurface. 
Figures~\ref{fig_std_gam1e2}c and  \ref{fig_std_gam1e2}d show $\dot K(s)$ and 
$\dot E(s)$ for the reference shell. Significant bending is confined to a boundary
layer of arcwise extent $\approx 0.14 R_0$ adjoining the edge of the shell. 
The inner (plateward) half of the boundary layer has $\dot K < 0$, which 
corresponds to counterclockwise bending. The outer half experiences
clockwise bending ($\dot K > 0$) due to the stresses applied by the outer
fluid flowing clockwise around the free edge of the slab. Turning to stretching,
we see that both stretching ($\dot E > 0$) and compression ($\dot E < 0$) are significant
in the boundary layer. Finally, the plate outside the boundary layer is under nearly 
uniform compression. 

Because the bending rate $\dot K$ and the stretching rate $\dot E$ have different units, the best way to compare them is
in terms of the rates of viscous dissipation of energy $\Phi$ associated with deformation by 
bending ($\Phi_b$) and stretching ($\Phi_s)$. For axisymmetric flow, the explicit 
expressions for these dissipation rates per unit area of the midsurface are
(\citet{novozhilov59}, Eq.~(9.12))
\be
\Phi_b = \frac{1}{6}\eta_1 h^3\left[ (\kappa_1 + \kappa_2)^2 - \kappa_1 \kappa_2\right],
\quad
\Phi_s = 2\eta_1 h \left[(\epsilon_1 + \epsilon_2)^2 - \epsilon_1\epsilon_2\right].
\ee
Figures~\ref{fig_std_gam1e2}e shows $\Phi_b(s)$ (red line) and $\Phi_s(s)$ (blue line)
for the reference shell. The inner half of the boundary layer is strongly 
dominated by bending, whereas the outer half shows a rough equipartition of 
dissipation between bending and stretching.

\subsection{Bending boundary layer}\label{sec:bbl}

\begin{figure}
	\centering
	\vspace{-6cm}
	\includegraphics[width=0.7\linewidth]{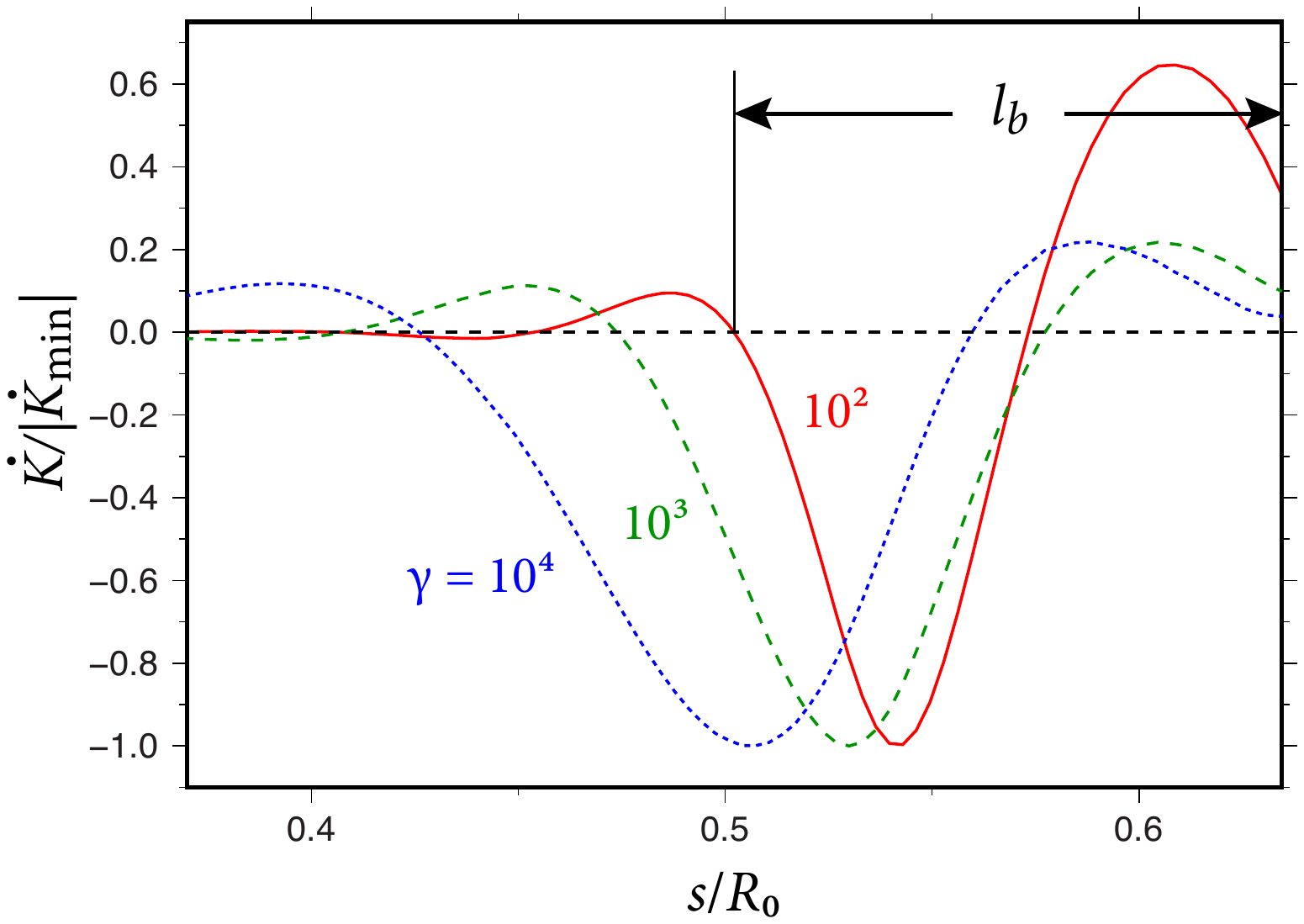}
	\vspace{-5cm}
	\caption{
		\label{fig_ellb_vs_gam}
		Bending rate $\dot K$ as a function of arclength for a shell with 
		$\theta_t = 30^{\circ}$, $\theta_s = 36^{\circ}$,
		$h/R_0 = 0.0157$, $d/h = 0.3$, $\varphi_s = 45^{\circ}$ and $\gamma = 10^2$ (red),
		$10^3$ (green) and $10^4$ (blue). The definition of the bending length $l_b$ is
		shown for $\gamma = 10^2$.
	}
\end{figure}

The next step is to define more precisely the width of the bending boundary layer,
which is just the bending length $l_b$ that we introduced in our scaling analysis
(\S~\ref{sec:physscaling}). 
Figure \ref{fig_ellb_vs_gam} shows the bending rate $\dot K(s)$ for the reference
shell geometry with viscosity ratios $\gamma = 10^2$ (red), $10^3$ (green) and 
$10^4$ (blue). As shown for the case $\gamma = 10^2$, we define $l_b$ as the distance
from the free end of the slab to the first zero of $\dot K$ plateward (leftward in the figure) from 
the position where $\dot K$ is a minimum. All else being equal, $l_b$ increases
as the viscosity ratio $\gamma$ increases: a stiffer shell bends over a greater distance. 

\begin{figure}
	\centering
	\vspace{-5cm}
	\includegraphics[width=0.7\linewidth]{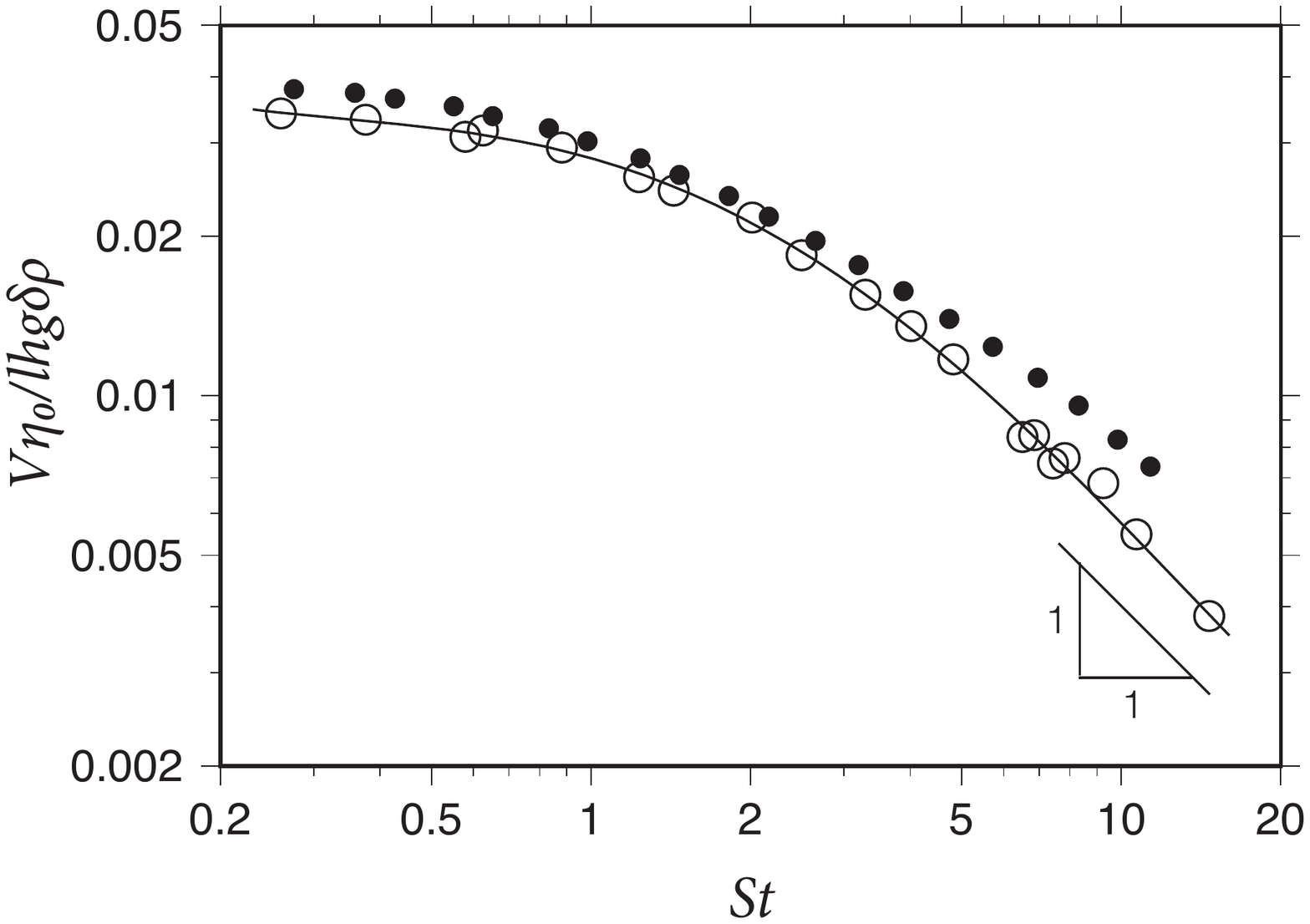}
	\vspace{-5cm}
	\caption{
		\label{fig_vvstokes_vs_st0}
		Normalised sinking speed
		$V/V_{\mathrm{Stokes}}$ as a function of the flexural stiffness $St$ for
		a slab with dip $\varphi_s = 45^{\circ}$. Open circles: predictions of 
		the axisymmetric BEM code in the flat-Earth limit $\theta_t = 90^{\circ}$, with $d/h = 0.3$ and various combinations of values of $\gamma$ and $l/h$.
		Solid line: polynomial fit to the axisymmetric BEM predictions. Filled circles: predictions of the two-dimensional Cartesian (``true flat Earth") BEM code of \citet{ribe10}.
	}
\end{figure}

To provide a basis of comparison for our subsequent solutions, we now characterise
in more detail the flat-Earth scaling law from Eq.~\eqref{vvstokesflat}. We 
performed BEM simulations with $h/R_0 = 0.0157$, $d/h = 0.3$, 
$\theta_t = 90^{\circ}$, $\varphi_s = 45^{\circ}$, $\theta_s - \theta_t\in [2^{\circ}, 4^{\circ}]$
and $\gamma\in [10^2, 10^{4.25}]$. The resulting values of $V/V_{Stokes}$ are shown
as a function of the flexural stiffness in figure~\ref{fig_vvstokes_vs_st0}. The points
collapse onto a universal curve with two limits. For $St\ll 1$, the slope of the 
curve approaches zero on logarithmic axes, indicating that the sinking speed is 
proportional to the Stokes speed Eq.~\eqref{vstokesdef} and therefore entirely controlled by
the exterior viscosity $\eta_0$. The operative balance here is between the external force $F_{ext}$ and the buoyancy force $F_{b}$ given by Eqs.~\eqref{fextscale} and \eqref{fbscale} respectively.
For $St > 10$, the slope is equal to $-1$, implying that the
sinking speed 
\be
V\sim \frac{g\delta\rho h l}{\eta_0}\left(\frac{\eta_1 h^3}{\eta_0 l_b^3}\right)^{-1} \propto \frac{1}{\eta_1}
\ee
depends only on the shell's own viscosity $\eta_1$. This behaviour
corresponds to a balance between the buoyancy force and the internal
viscous force $F_{int}$.

To verify that the hemispherical-plate limit $\theta_t = \pi/2$ does indeed correspond to flat-Earth behaviour, we show in figure~\ref{fig_vvstokes_vs_st0} the normalised sinking speed
$V(St)/V_{Stokes}$ predicted by the two-dimensional Cartesian BEM code of \citet{ribe10} (filled circles). Remarkably, the axisymmetric
and Cartesian predictions of $V(St)/V_{Stokes}$ agree to within
10-15\% for $St\leq 3$. This is all the more surprising given the
fact that the plate in the axisymmetric case is motionless while
the two-dimensional plate moves towards the trench.
Because $St<1$ is the relevant range of flexural stiffnesses for terrestrial subduction zones (see table \ref{tab:1}),
figure~\ref{fig_vvstokes_vs_st0} demonstrates that the case $\theta_t = \pi/2$ is a suitable proxy for subduction on a flat Earth.

\subsection{Effect of sphericity}

\begin{figure}
	\begin{center}
		\vspace*{0.0cm}
		\includegraphics[width=0.7\linewidth]{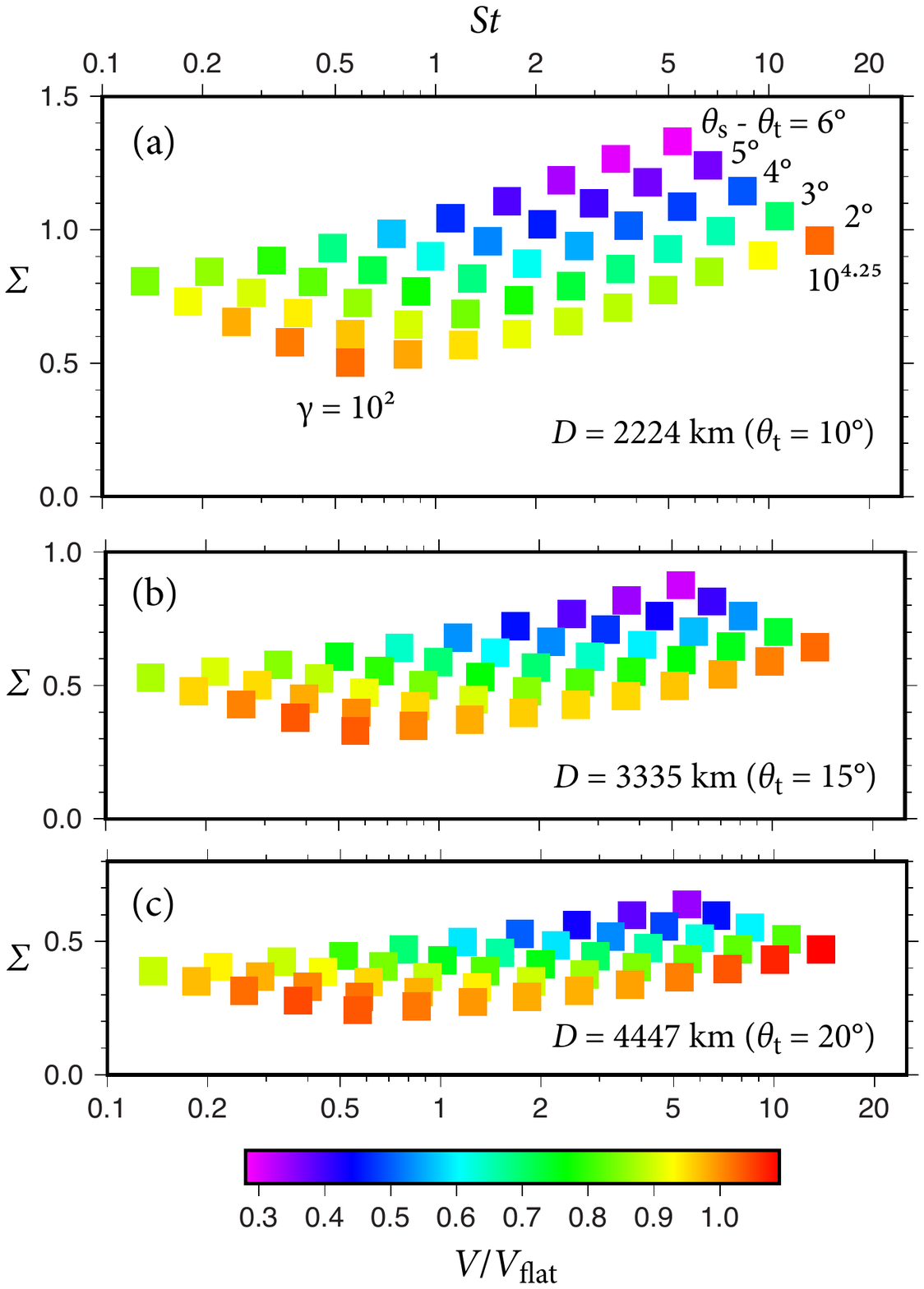}
	\end{center}
	\vspace*{-5cm}
	\caption{
		\label{fig_vvflat_vs_alpha}
		Slab sinking speed $V$ normalised to the flat-Earth limit (figure~\ref{fig_vvstokes_vs_st0})
		as a function
		of the flexural stiffness $St$ and the sphericity number $\Sigma$, for 
		three values of the arcwise diameter $D$ of the plate: (a) 2224 km, 
		(b) 3335 km and (c) 4447 km. Each panel comprises 50 numerical
		solutions with five different values of $\theta_s - \theta_t$ (labelled
		in part (a)) and ten viscosity ratios $\gamma\in [10^2, 10^{4.25}]$
		increasing from left to right. 
	}
\end{figure}

Turning now to solutions with sphericity ($\Sigma > 0$), we first calculate the ratio $V/V_{flat}$ of 
the sphericity-influenced sinking speed $V$ to the flat-Earth sinking speed $V_{flat}$.
The results are shown in figure~\ref{fig_vvflat_vs_alpha} as functions of $St$ and $\Sigma$
for three values of the arcwise diameter $D$ of the plate: (a) 2224 km ($\theta_t = 10^{\circ}$), 
(b) 3335 km ($\theta_t = 15^{\circ}$) and (c) 4447 km ($\theta_t = 20^{\circ}$).
Each panel is based on 50 BEM solutions with five different values of $\theta_s - \theta_t$
and ten different values of $\gamma$. In almost all cases $V/V_{flat} < 1$, indicating that
sphericity reduces the sinking speed of the slab by stiffening the shell. 
In general, sphericity has a greater influence for longer slab length, greater shell
stiffness $St$, and smaller plate diameter $D$. The last of these effects implies that the
influence of sphericity is strongest for plates in the form of shallow shells, i.e.\ shells whose
deviation from a plane is small compared to their radii of curvature. For the values of
$\theta_s - \theta_t$ and $\gamma$ used, the maximum influence of sphericity on the sinking speed
(in figure~\ref{fig_vvflat_vs_alpha}a) is nearly a factor of 4.

A feature of figure~\ref{fig_vvflat_vs_alpha} worth noting is that the
angular plate radius $\theta_t$ influences $V/V_{flat}$ in two ways: via its appearance in the definition of the sphericity number $\Sigma = (l_b/R_0)\cot\theta_t$,
and via its absolute value. Comparison of parts (a)-(c)
of figure~\ref{fig_vvflat_vs_alpha} immediately shows that 
$V/V_{flat}$ cannot be described as a function of $St$ and 
$\Sigma$ alone. This means that sphericity necessarily introduces
two dimensionless parameters ($\Sigma$ and $\theta_t$)
beyond those required to describe flat-Earth subduction,
as already anticipated in the scaling law Eq.~\eqref{vvstokesscale}.

Proceeding as we just did for the sinking speed, we now 
quantify the effect of sphericity on the longitudinal normal stress (`hoop stress') $\sigma_{22}$.
This quantity is important because compressional hoop 
stress is what causes longitudinal buckling of the slab. 
Although such buckling cannot occur in our axisymmetric model,
it nevertheless makes sense to calculate the axisymmetric
hoop stress as a basic state whose stability to longitudinal perturbations
can be analysed
(as we shall do in a subsequent study.)

\begin{figure}
	\begin{center}
		\vspace*{-10cm}
		\includegraphics[width=1.0\linewidth]{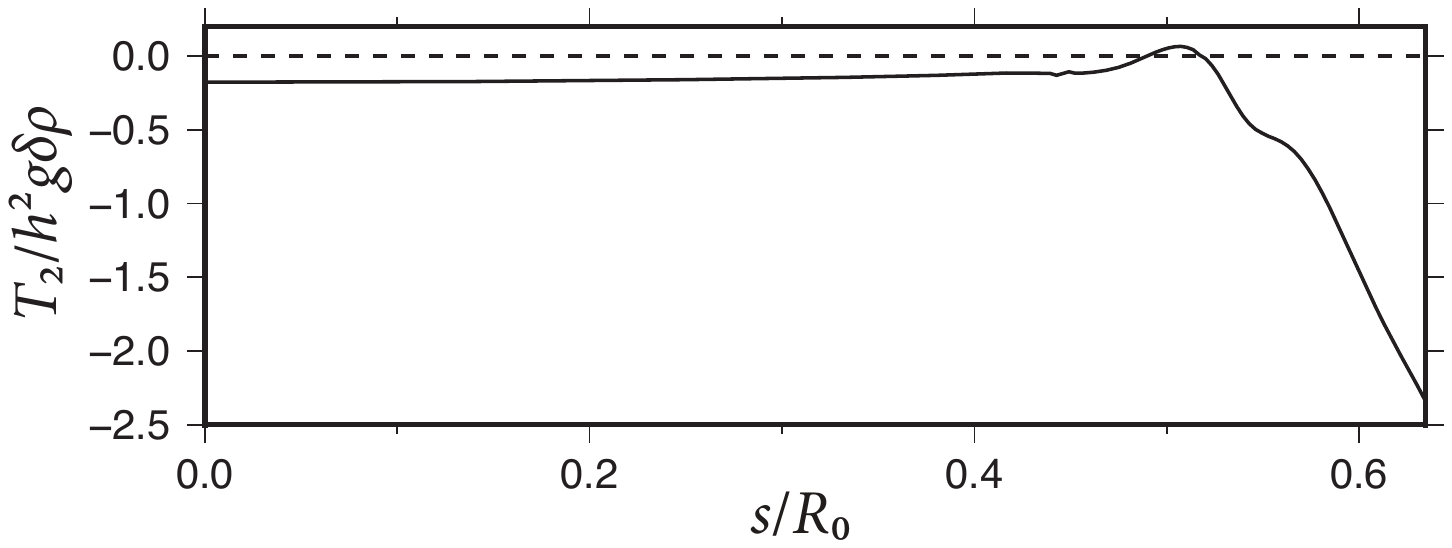}
	\end{center}
	\vspace*{-10cm}
	\caption{
		\label{fig_hoop_vs_s_std}
		Hoop stress resultant $T_2$ as a function of arclength for
		the same model parameters as in figure \ref{fig_std_gam1e2}.
	}
\end{figure}

In thin-shell theory \citep{novozhilov59}, the fundamental quantity
related to the hoop stress is the stress resultant
\be
T_2 = \int^{h/2}_{-h/2}\sigma_{22} (1 - K_1 z)\mathrm d z = 4\eta_1 h\left(\epsilon_2 + \frac{1}{2}\epsilon_1\right)
\label{hoopstress}
\ee
where $\epsilon_1$ and $\epsilon_2$ are defined by Eq.~\eqref{eps12}.
The notation $T_2$ is that of \citet{novozhilov59}; the single
subscript prevents confusion with the stress Green function $T_{ijk}$.
Figure \ref{fig_hoop_vs_s_std} shows $T_2$ as a function of 
arclength for the same model parameters as in 
figure \ref{fig_std_gam1e2}. Except for a small interval
near the trench, $T_2 < 0$ everywhere, indicating compressive
longitudinal stress. The maximum absolute value of $T_2$
occurs at the leading end of the slab.

\begin{figure}[t!]
	\begin{center}
		\vspace*{-5cm}
		\includegraphics[width=1.0\linewidth]{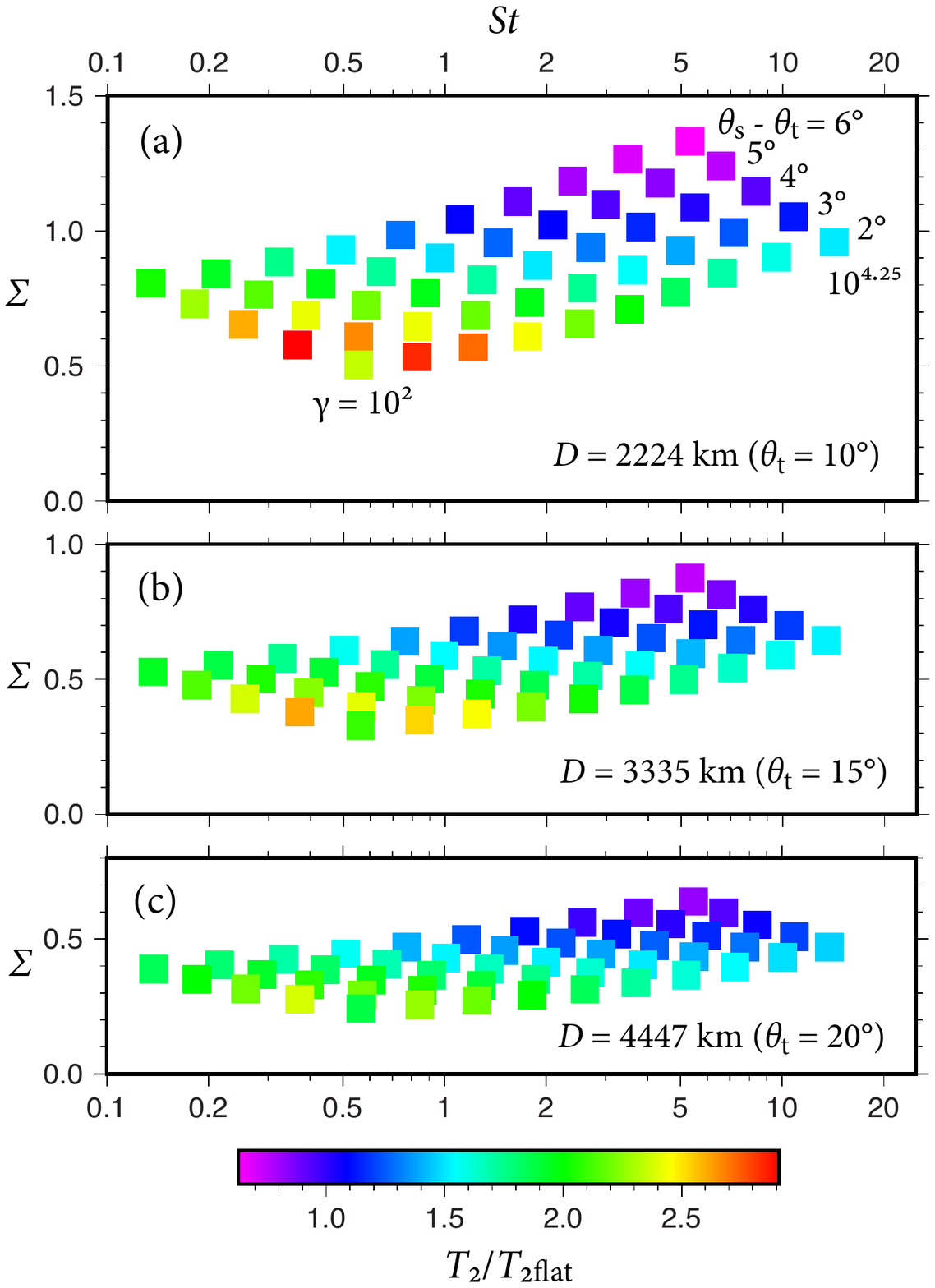}
	\end{center}
	\vspace*{-4cm}
	\caption{
		\label{fig_TTflat_vs_st_alpha}
		Same as figure \ref{fig_vvflat_vs_alpha}, but for
		the longitudinal normal stress resultant $T_2$ normalised
		to its flat-Earth value $T_{2flat}$.
	}
\end{figure}

Figure \ref{fig_TTflat_vs_st_alpha} shows $T_2$,
normalised to its flat-Earth value $T_{2flat}$, as a function
of the flexural stiffness $St$ and the sphericity number
$\Sigma$. Depending on the values of $St$ and $\Sigma$, 
$T_2/T_{2flat}$ can be either greater than or (not much) less than 
unity. The largest effect of sphericity in this case
occurs for relatively low values of $St$ and $\Sigma$.
This is opposite to the case of the normalised sinking
speed $V/V_{flat}$ (figure \ref{fig_vvflat_vs_alpha}), for
which the largest effect of sphericity occurs for 
relatively high values of $St$ and $\Sigma$.

\section{Illustrative time-dependent solutions}
\label{sec:timedep}

\begin{figure}
	\vspace*{-3cm}
	\begin{center}
		\includegraphics[width=0.8\linewidth]{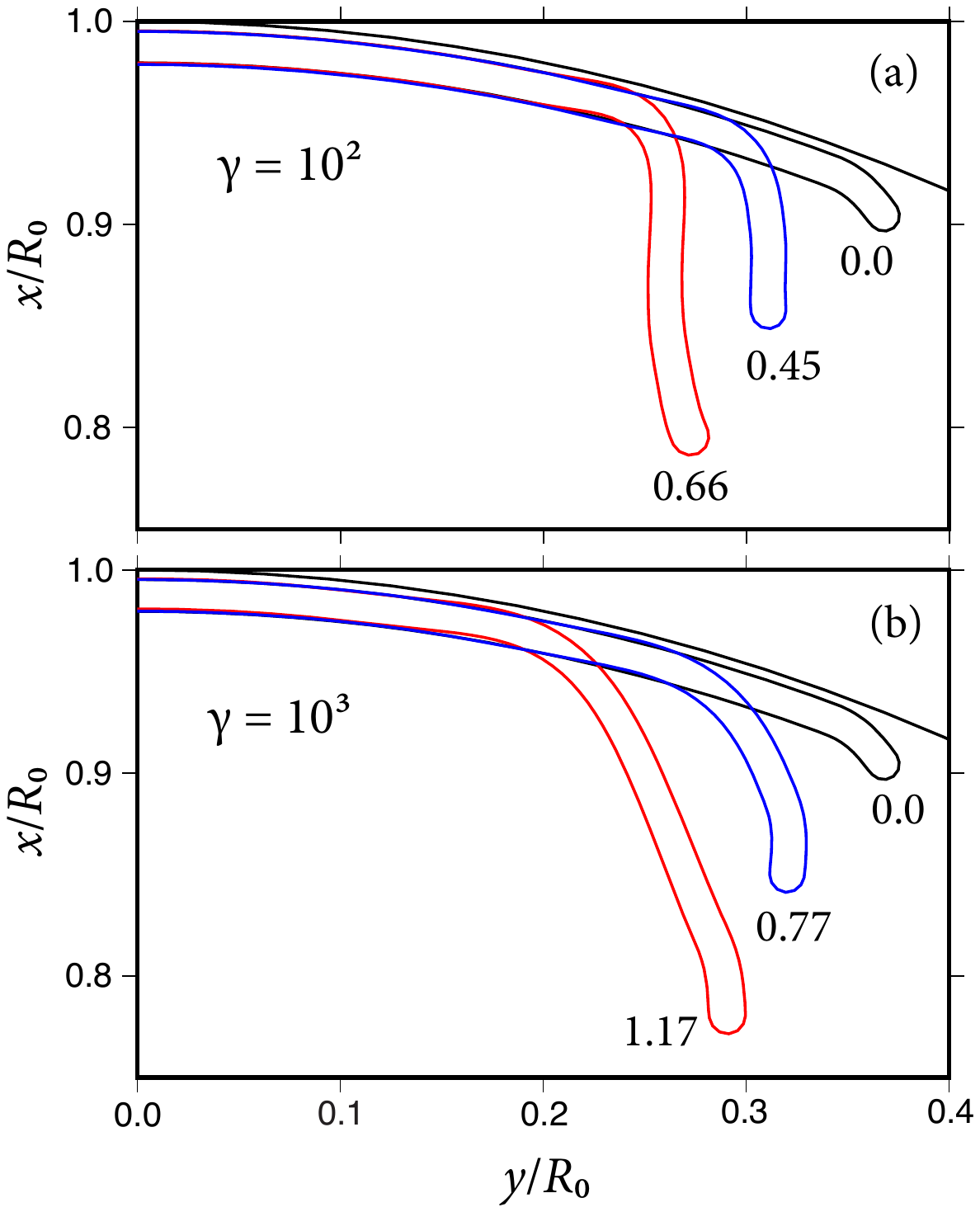}
	\end{center}
	\vspace*{-6cm}
	\caption{
		\label{fig_timedep}
		Time-dependent subduction of shells with viscosity ratios
		(a) $\gamma = 10^2$ and (b) $10^3$. The initial condition for both cases (black)
		has $\theta_t = 20^{\circ}$, $\theta_s = 22^{\circ}$, $\varphi_s = 30^{\circ}$,
		$h/R_0 = 0.0157$ and $d/h = 0.3$. Numbers near the end of each slab
		give the time in units of $R_0\eta_0/h^2 g\delta\rho$.
	}
\end{figure}

Having examined the scaling of instantaneous subduction dynamics
for fixed shell geometries, we now consider the time evolution of the shape
of the shell itself. As an illustration, figure~\ref{fig_timedep} shows the 
evolving shape of a shell for viscosity ratios $\gamma = 10^2$ and
$10^3$, starting from the initial condition shown in black. Numbers
near the end of each slab are dimensionless times $\hat t = t h^2 g\delta\rho/R_0\eta_0$. 
In both panels, the subducting shell gradually `peels away' from the 
free surface such that the trench undergoes retrograde (plateward) motion.
Geophysicists call this `trench rollback'. 
Comparing figures~\ref{fig_timedep}a and  \ref{fig_timedep}b, we see
that subduction evolves more rapidly for the lower viscosity 
ratio, simply because a less viscous shell can bend more easily. A further difference
between the two cases concerns the shape
of the slab. For $\gamma = 10^2$, the curvature of the lower part of the 
slab at $\hat t = 0.66$ is opposite in sign to that of the upper part. 
This is because the slab is weak enough to be deformed 
by the mantle `wind' flowing counterclockwise around its end. By
contrast, the curvature of the slabs with $\gamma = 10^3$ is of
the same sign everywhere because the slab is stiff enough
to resist deformation by the mantle wind. In closing, we
re-emphasise that these purely axisymmetric solutions will no doubt
be unstable to small longitudinal perturbations. 

\section{Geophysical application}
\label{sec:geophysical}

\begin{table}
	\centering
	\begin{tabular}{cccccccc}
		\hline
		Plate & Area (km$^2$)  & Subduction zone & $St$ & $\Sigma$ & $1 - V/V_{flat}$ & $T_2/T_{2flat}$ \\
		\hline
		PA & $1.05\times 10^8$ & Tonga & 0.13-0.37 & 0.12-0.13 & $\leq 0.065$ & 1.33-1.37 \\
		PA & $1.05\times 10^8$ & Marianas & 0.12-0.33 & 0.12-0.13 & $\leq 0.069$ & 1.45-1.64 \\
		NZ & $1.61\times 10^7$ & Chile & 0.062-0.17 & 0.48-0.52 & $\leq$ 0.20 & 1.29-1.44 \\
		PS & $5.44\times 10^6$ &  Ryukyu & 0.16-0.43 & 0.56-0.62 & 0.11-0.33 & 1.91-2.39 \\
		CO & $2.93\times 10^6$ & Central America & 0.091-0.25 & 0.66-0.72 & 0.11-0.34 & 1.91-2.31\\
		JF & $2.56\times 10^5$ & Cascadia & 0.029-0.083 & 2.3-2.5 & 0.12-0.33 & 1.69-2.19 \\
		\hline
	\end{tabular}
	\caption{Dimensionless parameters for Pacific subduction zones.}
	\label{tab:1}
\end{table}

\begin{figure}
	\vspace*{-4cm}
	\begin{center}
		\includegraphics[width=0.8\linewidth]{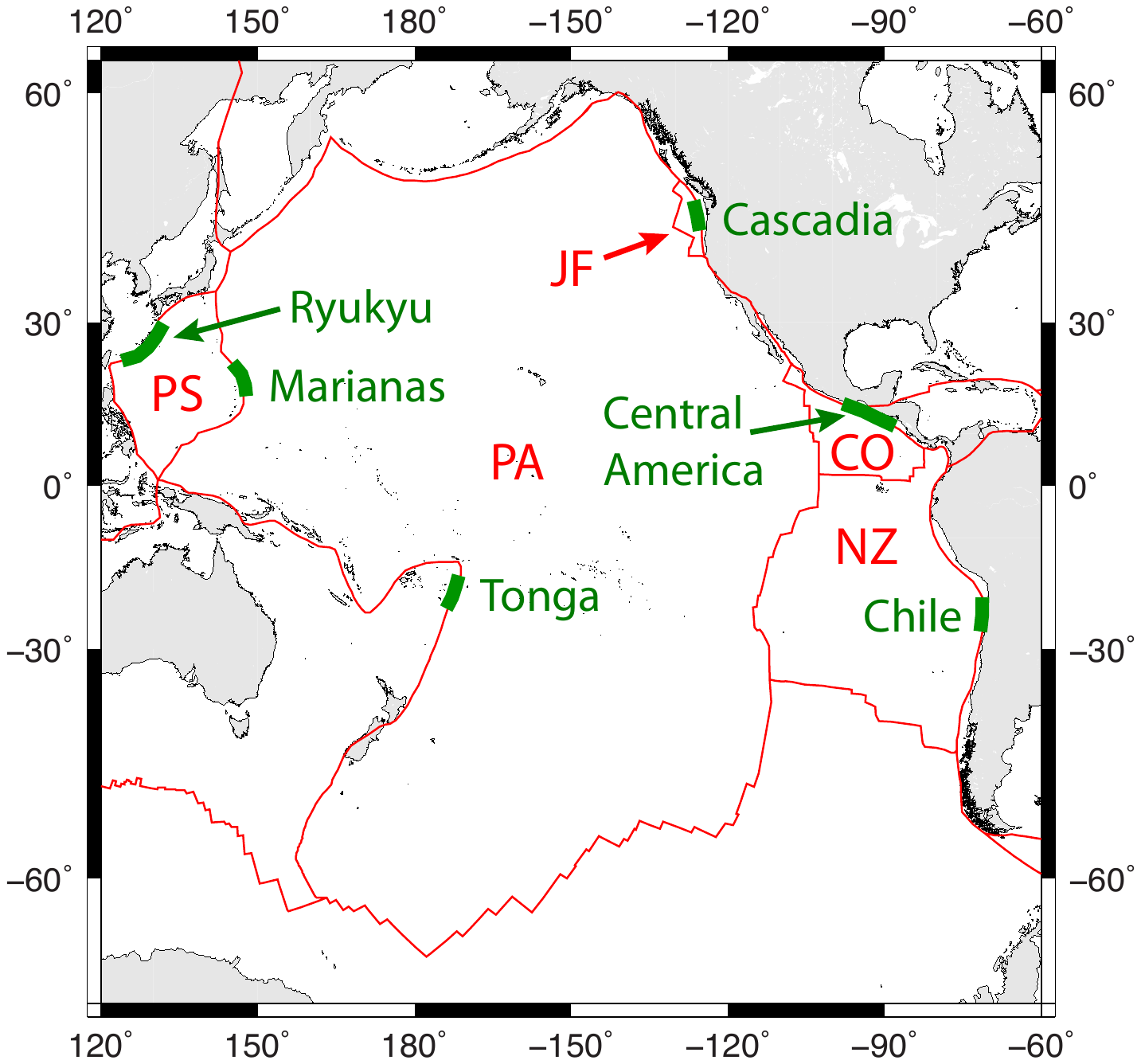}
	\end{center}
	\vspace*{-5cm}
	\caption{
		\label{fig_map}
		Map of the Pacific Ocean (Mercator projection) showing the boundaries of the major plates (red lines). Abbreviations of the plate names are shown in 
		red: CO = Cocos, JF = Juan de Fuca, NZ = Nazca, PA = Pacific, PS = Philippine Sea. The subduction zones that we consider are indicated by 
		green lines, and their names are shown in green. 
	}
\end{figure}

To apply our results to subduction on Earth, we now estimate the influence of 
sphericity on the dynamics of selected terrestrial subduction zones. Figure
\ref{fig_map} shows the locations (green) of the six subduction zones we have chosen, all located in the Pacific ocean basin: Tonga, Marianas (both involving subduction of the Pacific plate),
Ryukyu (Philippine Sea plate), Cascadia (Juan de Fuca plate),
Central America (Cocos plate), and Chile (Nazca plate). Because our model is axisymmetric,
we began by determining the radius $\theta_t$ of an equivalent spherical cap having the same area
as the plate in question. These areas are listed in column 2 of table \ref{tab:1}. Next, for each subduction zone we chose
the model parameters $h$, $\ell$, $\theta_s$ and $\varphi_s$ by averaging values given by 
\citet{lallemand05gcubed} for different transects perpendicular to the trench. 
We then ran instantaneous BEM simulations for the given slab geometry to determine
the sinking speed $V$ and the hoop stress resultant $T_2$ for ten viscosity ratios $\gamma\in [10^2, 10^{4.25}]$. 
Finally, we ran `flat-Earth' simulations for $\theta_t = 90^{\circ}$ and the same values of
$\theta_s - \theta_t$ and the other parameters to obtain ten values of the flat-Earth sinking speed $V_{flat}$ and the hoop stress resultant $T_{2flat}$. Appendix \ref{app:geophys} gives 
further details of these calculations.

\begin{figure}
	\vspace*{-5cm}
	\begin{center}
		\includegraphics[width=0.8\linewidth]{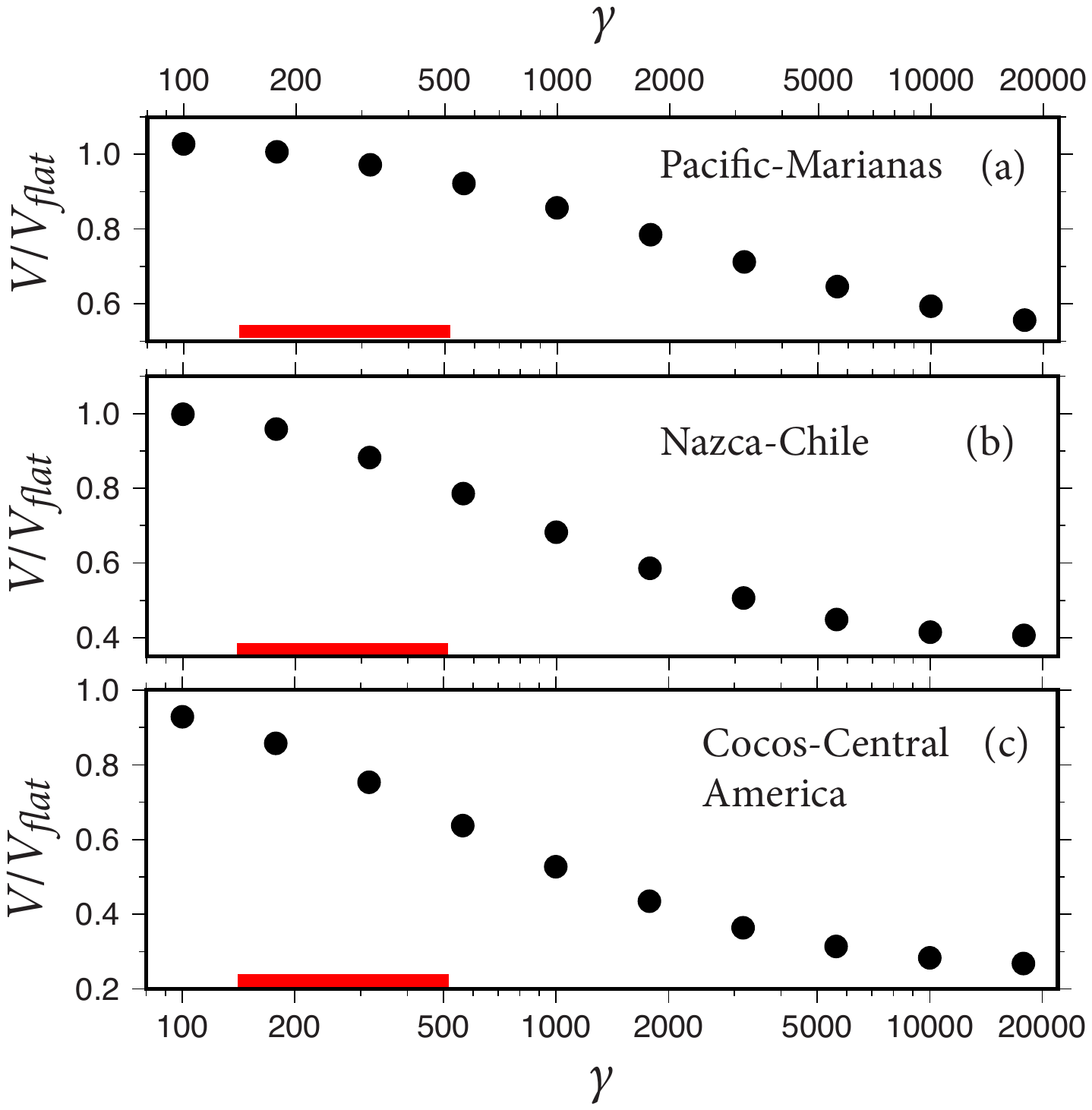}
	\end{center}
	\vspace*{-5cm}
	\caption{
		\label{fig_vvflat_vs_gam}
		Estimated effect of sphericity in selected terrestrial subduction zones
		as a function of the assumed plate/mantle viscosity ratio $\gamma$.
		(a) Marianas (Pacific plate),
		(b) Chile (Nazca plate), and (c) Central America (Cocos plate). Each panel shows the numerically 
		predicted slab sinking speed $V$ normalised to the flat-Earth limit as a function of the viscosity ratio. The red horizontal bars indicate the 
		plausible range of $\gamma$ for the Earth. 
	}
\end{figure}

Figure~\ref{fig_vvflat_vs_gam} shows $V/V_{flat}$ as a function of the viscosity
ratio $\gamma$ for three of our six subduction zones. Apart from one point in part (a) with 
$\gamma = 100$, $V/V_{flat} < 1$, indicating that sphericity decreases $V$ relative
to its flat-Earth value by stiffening the shell. In all cases, the effect of 
sphericity is stronger (smaller $V/V_{flat}$) for larger viscosity ratios. 
Finally, for a given viscosity ratio the sphericity effect is weak for
the Pacific plate, stronger for the Nazca plate, and strongest for
the Cocos plate. For the geometry of the Cocos plate, the sphericity
effect is nearly a factor of 4 for the highest viscosity ratio $\gamma = 10^{4.25}$.

To draw conclusions from figure~\ref{fig_vvflat_vs_gam} we need an estimate of
the effective viscosity ratio $\gamma$ for subducting slabs on Earth. Several lines of evidence
converge to suggest that $\gamma$ is on the order of a few hundred. \citet{houseman97gji} presented finite-element models of the Tonga slab and concluded that the modelled deformation
matched the observed if $\gamma\approx 200$. \citet{billen03gji} used
regional finite-element models of the Tonga-Kermadec subduction zone to conclude that 
the magnitude of the observed strain rate requires $\eta_1 < 3\times 10^{23}$ Pa.s 
or, equivalently, $\gamma < 300$. On the basis of analog laboratory
experiments, \citet{funiciello08} estimated that $\gamma\in [150,500]$ is required to 
explain the ratios of trench migration speeds to subducting plate speeds observed on Earth. 
An almost identical estimate $\gamma\in [140,510]$ was obtained by \citet{ribe10},
who compared observed minimum radii of curvature of subducted slabs with the predictions
of two-dimensional numerical models. \citet{capitanio12tectonophys} compared numerical
model predictions with the observed inverse correlation between slab dip and 
radius of curvature to conclude that $\gamma\approx 200$.

The red horizontal bars in figure~\ref{fig_vvflat_vs_gam} show the range of $\gamma$ 
that brackets the aforementioned estimates. The maximum effect of 
sphericity occurs at the high end of the range ($\gamma = 510$), and
is 
$1- V/V_{flat} = 7\%$ for Marianas, 20\% for Chile, and 
34\% for Central America. 

After estimating $1- V/V_{flat}$ for our chosen
subduction zones, we performed a 
similar calculation for the normalised longitudinal stress 
resultant $T_2/T_{2flat}$. The results of both calculations
are summarised in table \ref{tab:1} for all six subduction
zones and an assumed range of
viscosity ratios $\gamma\in [140, 510]$.
For the Tonga, Marianas and Chile 
subduction zones only the upper bound of $1- V/V_{flat}$
is given because calculation of the (small) lower bound is subject to
numerical error. Also shown are the corresponding ranges of
the flexural stiffness $St$ and the sphericity number $\Sigma$.
While $St$ is small ($\leq 0.43$) for all six subduction zones, 
$\Sigma$ shows a strong inverse correlation with the plate size,
as one would expect from the fact that $\Sigma\propto\cot\theta_t$.

\section{Discussion}
\label{sec:discussion}

We begin by discussing further some of the limitations of our
model, their motivations, and how they might be overcome. 
The first obvious limitation is the model's axisymmetry. One 
consequence of this is that the plate itself does 
not move, so that subduction occurs
by trench rollback alone. This situation is not encountered on Earth, 
where subducting plates usually have (crudely speaking) a mid-ocean ridge
on one side and a subduction zone on the other. 
The plate therefore can be pulled by the subducting
slab (`slab pull' in geophysical parlance), causing it to move from the ridge towards the 
subduction zone. Axisymmetry further implies that the
trench is everywhere convex inward. While this is the case 
on the large scale, a glance at figure \ref{fig_map} shows
that the subduction zones in the western Pacific ocean
are composed of adjoining concave-inward segments. However,
the purpose of the axisymmetry assumption is not geophysical
realism, but rather to formulate the simplest possible model
that allows us to quantify the influence of sphericity on
key dynamical aspects of subduction. Once the essential goal
of identifying the relevant length scales
and dimensionless parameters has been attained, one can proceed to investigate
three-dimensional subduction in more realistic geometries
that embody the features noted above. 

The second major simplification in our model is our neglect
of an effectively inviscid core, which on Earth has a radius $0.54 R_0$.
Such a core acts as a free-slip boundary condition
at the bottom of the mantle, and will have a significant
quantitative influence on the flow generated by subducting
slabs \citep{morra09, morra12}. We have two reasons for
ignoring the core. First, its presence would introduce an 
additional length scale into the model, 
making it harder to determine the underlying scalings while providing no
compensating physical insight. Second,
because no Green functions for a spherical annulus bounded
by two free-slip surfaces exist in the literature, a 
three-dimensional BEM approach would become unwieldy, requiring
discretisation of the entire surfaces of the Earth and of the core.
At this point one would probably be well advised 
to replace the BEM by more flexible approaches such as
finite element or finite volume methods, which allow additional
realistic aspects of the mantle (mineralogical phase
transitions, radial variation of viscosity, etc.) 
to be incorporated easily. As a final remark, we
emphasise that the quantities $V/V_{flat}$ and
$T_2/T_{2flat}$ that we calculate in this study are
normalised quantities with the flat-Earth limit
in the denominator. It is likely that the presence
of a core will influence the numerator and
the denominator similarly. If so, then our
estimates of the normalised quantities 
will remain valid even if a core is present. 

Our results show that the effect of sphericity depends strongly 
on the size of the subducting plate in a way that
defies naive intuition. At first glance, one might expect the
effect of sphericity to be greatest for a large plate, whose
midsurface differs more from a plane than that of a small or `shallow'
plate. However, the truth is exactly the opposite. Mathematically
speaking, this is so because the dimensionless number that 
measures the effect of sphericity is $\Sigma = (l_b/R_0)\cot\theta_t$. 
For the same value of $l_b$, therefore, $\Sigma$ is greater
for a small plate (small $\theta_t$) than for a large one. 
Physically, one can understand this result by cutting two
spherical shells from a child's plastic ball:
a large one equal to a full hemisphere, and a small one 
in the form of a shallow spherical cap. Now balance each shell
upside down on the point of an upright pencil, and 
deform the shell slightly by applying
radially directed forces to opposite sides. 
You will then see that a given applied force
produces a smaller deformation of the shallow cap than
of the hemisphere. In other words, the hemisphere
is `floppy' while the shallow cap is `stiff'. The
same result applies to viscous shells if one replaces
`deformation' by `rate of deformation'. 

From a geophysical point of view, the main conclusion of this
work is that sphericity affects the longitudinal
normal stress resultant $T_2$ much more than the
sinking speed $V$. This accords with the conclusion
of \citet{tanimoto98} that the dominant effect of sphericity 
is on the state of stress in subducted lithosphere.
We find that the effect of sphericity on $V$ is rather modest,
being at most about $7\%$ for the largest (Pacific) plate and
34\% for small plates (Philippine Sea, Cocos, Juan de Fuca) with $\theta_t\leq 12^{\circ}$. By contrast, the effect on $T_2$
is up to 64\% for the Pacific plate and up to 240\% for the small
plates. Our numerical solutions show that $T_2<0$ in the 
slab, which corresponds to compressional hoop stresses. These
stresses arise because a slab subducting in a sphere must
squeeze into a smaller and smaller space as it descends,
due to the diminution with depth of the surface area of 
concentric spheres. While this diminution could in principle
be accommodated by uniform thickening of the slab, it is 
more likely that the 
compressional hoop stresses drive longitudinal buckling instabilities
with growth rates enhanced by sphericity. The implication is that the
axisymmetric solutions we have discussed here will be
unstable to small non-axisymmetric perturbations. 
A BEM investigation of three-dimensional spherical subduction with dynamic buckling is
underway and will be reported separately.

Both authors thank E. Lauga for bringing us together. E. Calais kindly
provided a file of Earth's tectonic plate boundaries. We thank G. Morra and an anonymous referee for constructive comments that helped us significantly to improve the manuscript. N.M.R. was supported
by the Progamme National de Plan\'etologie (PNP) of the 
Institut des Sciences de l’Univers (INSU) of the CNRS, co-funded by CNES.

\section*{Declaration of Interests}

The authors report no conflict of interest.

\appendix

\section{Green functions}
\label{app:green}

In this Appendix we derive the Green's functions for the velocity and stress due to
a point force inside a fluid sphere with a constant viscosity $\eta$ and a free-slip outer
boundary condition. We consider separately the
cases of a radially-directed force and a transverse force with no radial component. Their application in our boundary element method is described in \S\ref{sec:bem}.

\subsection{Radial point force}

\begin{figure}
	\begin{center}
		\includegraphics[width=0.60\linewidth]{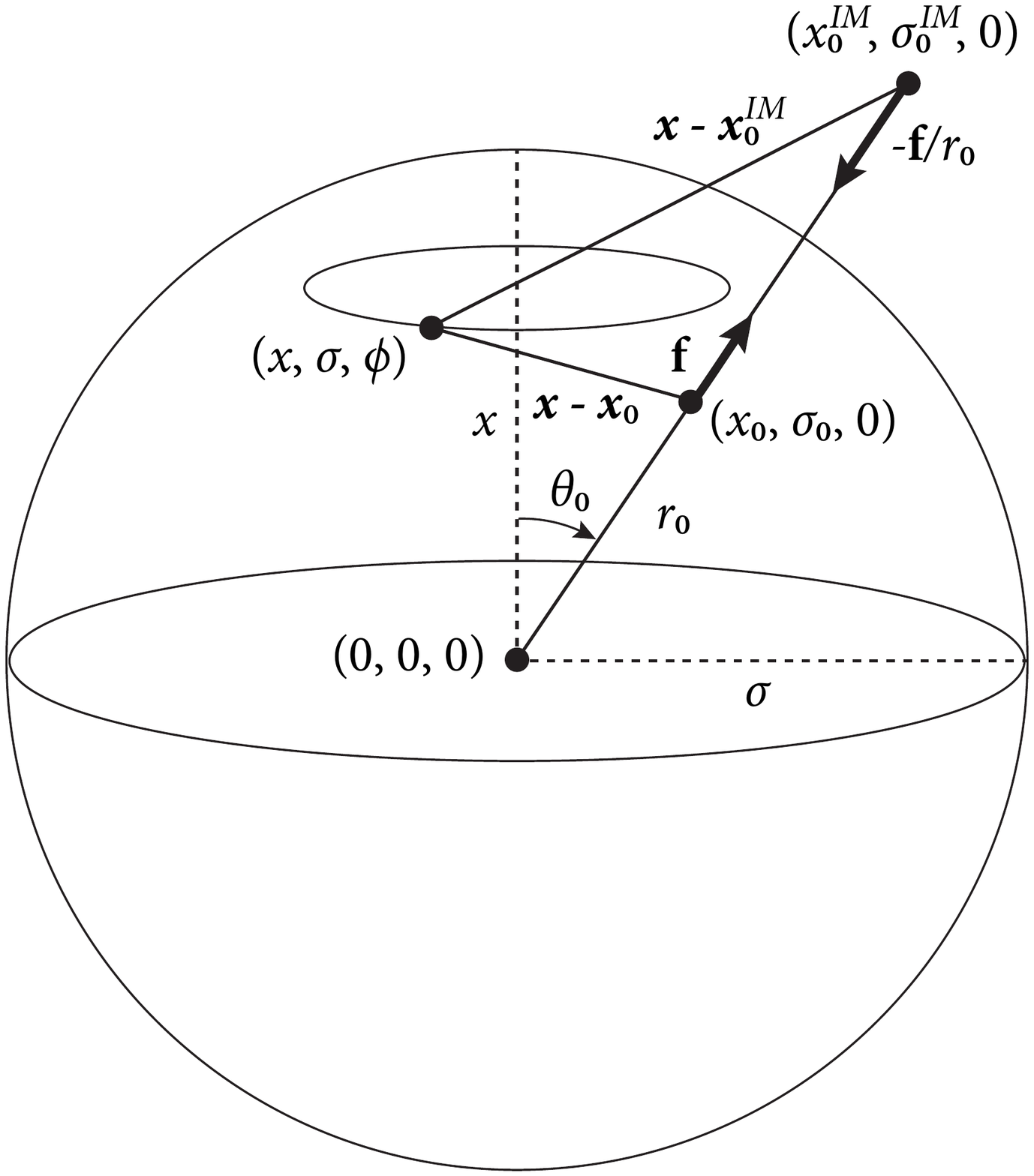}
	\end{center}
	\vspace*{-4cm}
	\caption{
		\label{fig_radialforce}
		Definition sketch for the image system of a radial point force
		$\bm{f}$ located at a field point $\bm{x}_0$ with cylindrical polar coordinates $(x_0, \sigma_0, 0)$.
		The colatitude of the field point is $\theta_0$ and its spherical radius is $r_0$. 
		The image $\bm{x}_0^{IM}$ of the point $\bm{x}_0$ outside the sphere 
		has cylindrical polar coordinates
		$(x_0^{IM}\equiv r_0^{-1}\cos\theta_0,\sigma_0^{IM}\equiv r_0^{-1}\sin\theta_0, 0)$.
		The force acting at the image point is $-r_0^{-1} \bm{f}$.
		An arbitrary integration point has cylindrical coordinates 
		$(x,\sigma, \phi)$ with the same values of $(r_0, \theta_0)$ as the field point.
	}
\end{figure}

The solution for this case is surprisingly simple, and can be obtained using
the method of images.
Figure~\ref{fig_radialforce} shows a sketch of the geometry.
Let $\bm{x}_0$ be the observation or field point with
colatitude $\theta$ and distance $r_0<1$ from the centre of the sphere.
The image of $\bm{x}_0$ outside the sphere is $\bm{x}_0^{IM} \equiv r_0^{-1} \bm{e}_r(\bm{x}_0)$.
Finally, $\bm{x}$ is another point  with the same colatitude $\theta$ and radius $r_0$
as $\bm{x}_0$; it will serve as the integration point when we perform an azimuthal integration
of the Green function.  
The cylindrical polar coordinates of $\bm{x}_0$ and 
$\bm{x}$ are $(x_0,\sigma_0,\phi_0)$ and $(x,\sigma,\phi)$,
respectively. Without loss of generality, we set $\phi_0=0$.

To satisfy the free-slip boundary conditions
on the sphere's surface $r = 1$, it suffices to add to the direct Stokeslet at $\bm{x}_0$ a second image Stokeslet 
of strength $-r_0^{-1} \bm{f}$
at the image point $\bm{x}_0^{IM}$. This is shown for the case $r_0>1$ in Eq.~(10.36) of \citet{kim91}, but continues to hold true for an internal singularity due to the symmetry of the solution. The radial Green function $\hat G_{ir}$ is therefore
\begin{align}
	\hat{G}_{ir}(\bx;\bx_0)&=J_{ij}(\bx-\bx_0)\left[\bm{e}_r(\bx_0)\right]_j-r_0^{-1}J_{ij}(\bx-\bx_0^{IM})\left[\bm{e}_r(\bx_0)\right]_j,	
	\label{hatgir}
\end{align}
where $\bm{e}_r(\bx_0)=\{\cos\theta_0,\sin\theta_0,0\}$ and
\be 
J_{ij}(\bm{r}) = \frac{1}{8\pi}\left(\frac{\delta_{ij}}{|\bm{r} |}
+ \frac{r_i r_j}{|\bm{r}|^3}
\right)
\ee
is the Oseen tensor. By analogy, the Green function
for the stress is 
\begin{align}
	\hat{T}_{irk}(\bx;\bx_0)&=K_{ijk}(\bx-\bx_0)\left[\bm{e}_r(\bx_0)\right]_j-r_0^{-1}K_{ijk}(\bx-\bx_0^{IM})\left[\bm{e}_r(\bx_0)\right]_j,
	\label{hattirk}
\end{align}
where
\be 
K_{ijk} (\bm{r}) = -\frac{3}{4\pi}
\frac{r_i r_j r_k}{|\bm{r}|^5}.
\ee

Given the expressions Eqs.~\eqref{hatgir} and \eqref{hattirk} for  $\hat G_{ir}$ and $\hat T_{irk}$ we must now
evaluate the integral expressions for the matrix elements $\hat{\mathcal M}_{\alpha\beta}$,
$\hat{\mathcal P}_{\alpha\beta\gamma}$ and $\hat{\mathcal Q}_{\alpha\beta\gamma}$.
To do this, we write the components of
$\bm{\hat{G}}$ and $\bm{\hat{T}}$ in terms of cylindrical coordinates by setting
$x - x_0 = \hat x$, 
$y - y_0 = \sigma\cos\phi - \sigma_0\cos\phi_0$, 
$z - z_0 = \sigma\sin\phi - \sigma_0\sin\phi_0$, 
$x - x_{IM} = \hat x_{IM}$, 
$y - y_{IM} = \sigma\cos\phi - \sigma_{IM}\cos\phi_0$, 
and
$z - z_{IM}  = \sigma\sin\phi - \sigma_{IM}\sin\phi_0$.
In the foregoing expressions, the angle $\phi_0$ has been retained for clarity, even though $\phi_0 = 0$ because
we are working in the $x$-$y$ plane by hypothesis.
While it is possible to write down closed-form expressions for $\hat{\mathcal M}_{\alpha\beta}$,
$\hat{\mathcal P}_{\alpha\beta\gamma}$ and $\hat{\mathcal Q}_{\alpha\beta\gamma}$ in terms of 
complete elliptic integrals, the
results are rather complicated. Accordingly, 
we chose instead to evaluate the integrals numerically. This also provides for greater consistency with
the case of a transverse point force (next subsection), for which one has no choice but to perform the azimuthal integrals
numerically. 

\subsection{Transverse point force}
\label{sec:transforce}

The Green's functions for this case can be obtained by adapting the solution of 
\citet{padmavathi95fdr} for flows inside an impermeable spherical boundary enclosing singularities 
and separating fluids having different viscosities. The limit relevant to our problem is 
when the outer viscosity is zero, corresponding to a free-slip surface. 

To this end, we write the solution of the Stokes equations 
in terms of two scalar potentials $\left\{A,B\right\}$, as
\begin{subequations}
	\begin{align}
		\bm{u}&=\bn\times\left(\bn\times\left(A\bx\right)\right) + \bn\times\left(B\bx\right),\\
		p&=p_0+\eta\frac{\partial}{\partial r}\left(r\nabla^2A\right),
	\end{align}
	\label{upab}
\end{subequations}
where $r=\sqrt{x^2+y^2+z^2}$ and the potentials satisfy
\begin{align}
	\nabla^4A=0,\quad\nabla^2B=0.
\end{align}
The condition of impermeability of the surface $r=1$ is satisfied if 
\begin{align}\label{eq:BCnopen}
	A=0\quad\text{on }r=1.
\end{align}
The no-stress boundary condition, $\sigma_{r\theta}=\sigma_{r\phi}=0$ is satisfied when
\begin{align}\label{eq:BCnostress}
	\frac{\partial^2 A}{\partial r^2}=0,\quad \frac{\partial}{\partial r}\left(\frac{B}{r}\right)=0\quad\text{on }r=1.
\end{align}
We therefore have a total of three boundary conditions that the two scalar fields $A$ and $B$ must satisfy.

The potentials for a transverse Stokeslet with strength $f_\theta\bm{e}_{\theta,0}$ are 
\begin{subequations}
	\begin{align}
		A^0(\bx;\bx_0)&=\frac{f_\theta}{8\pi\eta}\left(\frac{\bm{x}\cdot\bm{e}_{r,0}-r_0}{r_0}\left(R-r\right)+r-r_0\frac{\bm{x}\cdot\bm{e}_{r,0}}{r}\right)\frac{\bm{x}\cdot\bm{e}_{\theta,0}}{|\bm{x}\times\bm{e}_{r,0}|^2},\\
		B^0(\bx;\bx_0)&=\frac{f_\theta}{4\pi\eta }\left(\frac{R-r}{r_0}+\frac{\bm{x}\cdot\bm{e}_{r,0}}{r}\right)\frac{\bm{x}\cdot\bm{e}_{\phi,0}}{|\bm{x}\times\bm{e}_{r,0}|^2}.\label{eq:tvstokesletB}
	\end{align}
	\label{abfreespace}
\end{subequations}
Here the notation $\bm{e}_{r,0}=\bm{e}_r(\bx_0)$ was introduced for brevity.
The superscript $^0$ denotes a free-space potential, $(\bx;\bx_0)$ denotes the value of the potential at position $\bx$ due to the singularity at $\bx_0$, $r=|\bx|$ and $R=|\bx-\bx_0|$ is the distance between the two points. For clarity, we use two ways of expressing coordinate dependence, $(\bx;\bx_0)$ and $(r,\theta,\phi;r_0,\theta_0)$, interchangeably as needed.

\citet{padmavathi95fdr} proved a theorem that makes it possible to construct potentials that satisfy a certain set of boundary conditions on the unit sphere from the free-space Green functions Eq.~\eqref{abfreespace}. For the no-shear boundary conditions Eq.~\eqref{eq:BCnopen} and Eq.~\eqref{eq:BCnostress} and potentials $\{A^0, B^0\}$ that are singular only within the ball $r=1$, the solution for $r<1$ is given by
\begin{subequations}\label{eq:spheretheorem}
	\begin{align}
		A(r)&=A^0\left(r\right)-rA^0\left(\frac{1}{r}\right)\\
		B(r)&=B^0\left(r\right)+\frac{1}{r}B^0\left(\frac{1}{r}\right)-3r\int_{\infty}^{\frac{1}{r}}\xi B^0\left(\xi\right)\, \text{d}\xi,
	\end{align}
\end{subequations}
where we have omitted the dependence on the variables $\{\theta,\phi;r_0,\theta_0\}$ for brevity. 
In order for the integral to be well-defined, it is necessary that $B^0(r)$ decays faster than $r^{-2}$ as $r\to\infty$. 
However, as can be seen from an expansion of Eq.~\eqref{eq:tvstokesletB}, this is not the case for the transverse Stokeslet.
In fact, the Green's function for a single transverse force does not exist, because it would exert 
a net torque on the fluid that cannot be balanced due to the no-shear boundary condition on the sphere surface. 
Following \citet{padmavathi95fdr}, we resolve this problem by adding a rotlet singularity 
with strength $-f_\theta r_0\bm{e}_{\phi,0}$ at the centre of the sphere to balance the torque associated
with the transverse Stokeslet. 
The potentials for the rotlet are 
\begin{subequations}
	\begin{align}
		A^0(\bx;\bx_0)&=0,\\
		B^0(\bx;\bx_0)&=-\frac{f_\theta r_0}{8\pi\eta}\frac{\bm{x}\cdot\bm{e}_{\phi,0}}{r^3}.
	\end{align}
\end{subequations}
The combined potentials for a transverse Green function are then
\begin{subequations}\label{eq:tvstokeslet}
	\begin{align}
		A^0(\bx;\bx_0)&=\frac{f_\theta}{8\pi\eta}\left(\frac{\bm{x}\cdot\bm{e}_{r,0}-r_0}{r_0}\left(R-r\right)+r-r_0\frac{\bm{x}\cdot\bm{e}_{r,0}}{r}\right)\frac{\bm{x}\cdot\bm{e}_{\theta,0}}{|\bm{x}\times\bm{e}_{r,0}|^2},\\
		B^0(\bx;\bx_0)&=\frac{f_\theta}{4\pi\eta }\left(\frac{R-r}{r_0}+\frac{\bm{x}\cdot\bm{e}_{r,0}}{r}-r_0\frac{|\bm{x}\times\bm{e}_{r,0}|^2}{2r^3}\right)\frac{\bm{x}\cdot\bm{e}_{\phi,0}}{|\bm{x}\times\bm{e}_{r,0}|^2}
	\end{align}
	\label{a0b0}
\end{subequations}
The potentials in Eq.~\eqref{eq:tvstokeslet} can now be injected into Eq.~\eqref{eq:spheretheorem} to obtain a solution that satisfies the boundary conditions at $r=1$, and that behaves asymptotically like a Stokeslet near $\bm{x}_0$ and like a rotlet near $\bm{0}$. In order to employ this Green's function we additionally need to demonstrate that the non-local rotlet singularity does not cause a problem in the boundary-element formalism, which we do in Appendix \ref{app:justif}. 

With expressions for the potentials $A$ and $B$ in hand,
we can now use Eq.~\eqref{upab} to determine the velocities, pressure, and stress tensor associated with a unit transverse force.
In order to keep the algebra remotely manageable, we consider a ``standard" unit point force located on the 
$x$-axis at $(r_0, 0,0)$ and pointing in the $y$-direction, and use rotations to generalise this to arbitrary positions and orientations. Let $(u^s_x, u^s_y, u^s_z)$ be the Cartesian components of the velocity due to this point force,
and let $p^s$ be the pressure. Furthermore, define the auxiliary quantities
\be
c_1 = \sqrt{r^2 + r_0^2 - 2 r_0 x},\quad
c_2 = \sqrt{1 + r^2 r_0^2 - 2 r_0 x}.
\ee
Then we find
\begin{align}
	u^s_x =& \frac{y}{8\pi}\left[ \frac{x}{c_1^3} + \frac{r_0^2 x}{c_2^3} + r_0\left(\frac{5}{2} - \frac{1}{c_1^3} - \frac{1}{c_2^3} - \frac{3}{c_2} + \frac{1}{r^3} - \ln 8 \right) + 3 r_0\ln (1 + c_2 - r_0 x)\right],\label{usx} \\
	u^s_y =& \frac{1}{16\pi}\left\{ \frac{2 (2 + r^2 r_0^2 - 3 r_0 x)}{c_2^3 (r_0 x - 1)} + \frac{4 r^2 + 4 r_0^2 - 8 r_0 x - 2 (x^2 + z^2)}{c_1^3} \right. \nonumber\\
	& + r_0^2\left[ -\frac{2}{c_1^3} + \frac{2 (4 + 3 r^2 r_0^2 - 7 r_0 x)(x^2 + z^2)}{c_2^3 (r_0 x - 1)} \right]\nonumber \\
	& + r_0 x\left[ \ln 64 - 5 + \frac{4}{c_1^3} + \frac{4}{c_2^3} - \frac{2}{r^3} \right. \nonumber \\
	& \left. \left. -\frac{8}{c_2 (r_0 x - 1)}  + \frac{6 (c_2 + r_0 x - 1) z^2}{(r_0 x - 1)(x^2 - r^2)}\right]- 6r_0 x\ln (1 + c_2 - r_0 x)\right\}, \label{usy}\\
	u^s_z =& \frac{1}{8\pi}\left\{ \frac{1}{c_1^3} + \frac {r_0^2\left[4 + 3 r^2 r_0^2 - 7 r_0 x + c_2 (4 + 3 r^2 r_0^2 - 6 r_0 x)\right]}{c_2^3 (1 + c_2 - r_0 x)} \right\} y z, \label{usz}\\
	p^s =& \eta \frac{y\left[  c_2^3 (r^2 + 3 c_1^3 r_0 x - x^2) + c_1^3 (4 r^2 r_0^2 + 3 r^4 r_0^4 - 3 r_0 x - 9 r^2 r_0^3 x + 5 r_0^2 x^2)\right]}{4\pi c_1^3 c_2^3 (y^2 + z^2)} \label{ps}
\end{align}

The corresponding expressions for the components of the strain-rate tensor $e^s_{ij}$
are not written down here due to their length, but are readily determined by differentiating
Eqs.~\eqref{usx}-\eqref{usz}. The stress tensor is then
$\sigma^s_{ij} = - p^s\delta_{ij} + 2 \eta e^s_{ij}$ as usual.

Finally, for a general transverse force located at position $(x,y,z) = (r_0\cos\theta_0, r_0\sin\theta_0, 0)$, we obtain the 
desired tensor components $\hat G_{i\theta}$ and $\hat T_{i\theta k}$ as
\begin{align}
	\hat G_{i\theta} &=  u^s_j(\bm{R}\cdot\bm{x}) R_{ji} , \label{ghattrans} \\
	\hat T_{i\theta k} &= \eta^{-1}R_{ji} \sigma^s_{jl}(\bm{R}\cdot\bm{x}) R_{lk }, \label{hatttrans}
\end{align}
where
\be
\bm{R} = 
\begin{pmatrix}
	\cos\theta_0 & \sin\theta_0 & 0\\
	-\sin\theta_0 & \cos\theta_0 & 0 \\
	0  &  0 &  1\\
\end{pmatrix}, 
\ee
is a rotation matrix. In Eqs.~\eqref{ghattrans} and \eqref{hatttrans} the quantities $u^s_i$ and $\sigma^s_{ij}$ 
are evaluated at $\bm{R}\cdot\bm{x}$.

\section{Justification of the boundary element equation with a non-local singularity}
\label{app:justif}

In this section we derive the boundary integral equation for our numerical scheme. Since this is largely a standard calculation we focus here on the effect of the rotlet singularity at the origin and refer the reader to \citet{pozrikidis92} for details.

We define the velocity and stress Green's functions $G_{ij}(\bx;\bx_0)$ and $T_{ijk}(\bx;\bx_0)$ in analogy with Pozrikidis's notation. As $\bx\to\bx_0$ these behave asymptotically like a Stokeslet as usual, while as $\bm{x}\to\bm{0}$ we have
\begin{align}
	G_{ij}\sim\Pi_{jm}\frac{\varepsilon_{iml}x_l}{r^3},\quad T_{ijk}\sim-3\Pi_{jm}\frac{\varepsilon_{iml}x_kx_l+\varepsilon_{kml}x_ix_l}{r^5},
\end{align}
where $\Pi_{jm}(\bx_0)$ is a  projection operator that ensures that the rotlet is only present for transverse forcing. These diverge as $r^{-2}$ and $r^{-3}$ respectively. Hence in order to satisfy the divergence theorem, Eq.~(2.3.7) in \citet{pozrikidis92} needs to be augmented by an additional integral over the surface $S_0$ of a small ball surrounding the origin. As a result we have
\begin{align}\label{eq:modBEM}
	&8\pi\eta u_j(\bx_0)+\int_D \left[G_{ij}(\bx;\bx_0)\sigma_{ik}(\bx)-\eta u_i(\bx)T_{ijk}(\bx;\bx_0)\right]n_k(\bx)\,\text{d}S(\bx)\nonumber\\
	=&-\Pi_{jm}(\bx_0)\int_{S_0}\left[\frac{\varepsilon_{iml}n_l}{r^2}\sigma_{ik}(\bx)+3\eta u_i(\bx)\frac{\varepsilon_{iml}n_kn_l+\varepsilon_{kml}n_in_l}{r^3}\right]n_k r^2\,\text{d}\Omega,
\end{align}
where $n_k=x_k/r$ and we recover the boundary integral formalism if the RHS vanishes. The first term in the integral on the RHS tends smoothly to
\begin{align}
	\sigma_{ik}(\bm{0})\int_{S_0}\varepsilon_{iml}n_ln_k\,\text{d}\Omega\propto\sigma_{ik}\varepsilon_{imk}=0,
\end{align}
by symmetry of the stress tensor. Meanwhile, the second term in the integrand diverges, so we need to Taylor-expand $u_i(\bx)$ about the origin and use the identity  $\varepsilon_{kml}n_kn_l=0$ to find
\begin{align}
	&\int_{S_0}3\eta u_i(\bx)\frac{\varepsilon_{iml}n_l}{r}\,\text{d}\Omega\nonumber\\
	&=\frac{3\eta u_i(\bm{0})}{r}\varepsilon_{iml}\int_{S_0}n_l\,\text{d}\Omega+3\eta\int_{S_0}\varepsilon_{iml}n_ln_b\partial_bu_i(\bm{0})\,\text{d}\Omega+O\left(r\right)\nonumber\\
	&\rightarrow4\pi\eta\omega_m(\bm{0}),
\end{align}
since $\int_{S_0}n_l\,\text{d}\Omega=0$. Here $\bm{\omega}=\bn\times\bu$ is the vorticity of the flow. However, since the setup and flows we consider in this paper are axisymmetric, the vorticity at the origin is zero by construction. In conclusion, Eq.~\eqref{eq:modBEM} reduces to the standard boundary integral equation and therefore we do not need to worry about any numerical consequences due to the additional singularity in the Green function. Physically this is because there is no net torque on the sphere for axisymmetric motion, and hence the singularities at the origin due to locally non-zero tangential forces cancel out.

\section{Test problem: Concentric viscous spheres}
\label{app:concentric}

Our goal is to develop a simple analytical flow solution that
can be used to verify the correctness of our boundary-element code. We consider the
flow due to a buoyant fluid sphere of radius $R_1$ rising/sinking inside a concentric 
sphere of radius $R_0$ with a free-slip surface
containing fluid with a different viscosity. The fluid between
the two spherical surfaces has viscosity $\eta_0$ and density $\rho$, and
the fluid inside the inner sphere has viscosity $\eta_1\equiv \gamma\eta_0$
and density $\rho + \delta\rho$. Gravity is directed along the Cartesian
$x$-axis, i.e.\ it is not radial as it is in the Earth. While \citet{happel91}
solve a number of similar problems, the case of a free-slip outer sphere
is not among them.

In the equations discussed below, lengths are non-dimensionalised
by $R_0$ and velocities by $g\delta\rho R_0^2/\eta_0$, where $g$ is the 
gravitational acceleration.

The flow in both fluids can be described by poloidal scalars
$\Phi_i (r,\theta)$, where $i = 0$ or 1 is the fluid index. The velocity
in each fluid is defined in terms of the poloidal scalars as
\be
\bm{u}_i = - \frac{\bm{e}_{\theta}}{r} \partial^2_{r\theta}\left(r\Phi_i\right)
+ \frac{\bm{e}_r}{r}\mathcal B^2\Phi_i,
\quad
\mathcal B^2 = \frac{1}{\sin\theta}\partial_{\theta}\left(\sin\theta\,\partial_{\theta}\right)
\ee
where $\bm{e}_{\theta}$ and $\bm{e}_r$ are unit vectors in the indicated directions.
The poloidal scalars satisfy the biharmonic equation
\be
\nabla^4\Phi_i = 0.
\label{biharm}
\ee
Previous experience \citep{happel91} shows that $\Phi_i\propto\cos\theta$
for this problem. The corresponding general solution is 
\be
\Phi_i = \cos\theta
\left(
a_i r + b_i r^{-2} + c_i r^3 + d_i
\right)
\label{phi1soln}
\ee
The pressure
associated with Eq.~\eqref{phi1soln} is
\be
p_i = - 2\eta_i (10 c_i r + d_i r^{-2})\cos\theta
\label{press}
\ee
where $\eta_0= 1$ and $\eta_1 = \gamma$.
To determine the eight constants $a_0$-$d_1$, we first require the 
solution to be finite at $r=0$, which implies $b_1= d_1 = 0$. 
Next we apply the free-slip conditions $\Phi_1(1,\theta) = \partial^2_{rr}\Phi_1(1,\theta) = 0$
on the outer surface $r=1$.
Finally, we apply four matching conditions on the normal and tangential components
of the velocity and stress at the interface $r=\beta$. The inhomogeneous condition
that drives the flow is the matching condition on the normal stress $\sigma_{rr}$, which has the form
\be
(\sigma_{rr})_0(\beta,\theta) - (\sigma_{rr})_1(\beta,\theta) = x(\beta,\theta)\equiv \beta\cos\theta.
\ee
The explicit expressions for the constants are
\be
b_1 = d_1 = 0, 
\quad 
d_0 = - a_0 = \frac{\beta^3}{6},
\ee
\be
a_1 = -\frac
{\beta^2\left[-3 + \beta^5 (2\beta - 3) (\gamma-1) -2\gamma + \beta (2 + 3\gamma)\right]}
{D(\beta,\gamma)},
\ee
\be
c_0 = -b_0 = \frac{\beta^5\gamma}{D(\beta,\gamma)},
\quad
c_1 = \frac{\beta^5 - 1}{D(\beta,\gamma)},
\ee
where 
\be
D(\beta,\gamma) = 6 \left[2 + 2\beta^5 (\gamma -1 ) + 3\gamma\right].
\ee
For testing purposes, the main parameter of interest is the ascent/descent speed $U$ 
of the inner sphere. This is given by the radial (=vertical)
velocity at the North pole $(r,\theta) = (\beta,0)$ of the inner sphere, or
\be
\frac{U\eta_0}{g\delta\rho R_0^2} = \frac
{\beta^2\left[ 2\beta^6 (\gamma-1) - \beta^5 (3\gamma-2) - 2(\gamma + 1) + \beta (3\gamma + 2)\right]}
{3\left[ 2 + 2\beta^5 (\gamma - 1) + 3\gamma\right]}\equiv G(\beta,\gamma)
\label{uform}
\ee
where $U$ is dimensional. To verify the correctness of Eq.~\eqref{uform}, we consider the 
limit $\beta\to 0$ corresponding to a very small inner sphere, and re-dimensionalise $U$ using the 
radius $R_1$ of that sphere. We thereby find
\be
\frac{U\eta_0}{g\delta\rho R_1^2} \equiv \lim_{\beta\to 0}  \beta^{-2} G(\beta,\gamma)
= - \frac{2 (\gamma + 1)}{3 (3\gamma + 2)}.
\label{risespeed}
\ee
The expression Eq.~\eqref{risespeed} agrees exactly with the Stokes-Hadamard-Rybczynski
solution for a fluid sphere moving in an infinite fluid with a different viscosity.

\section{Calculations for Pacific subduction zones}
\label{app:geophys}

The first step in the calculations is to determine the angular radius 
\be
\theta_t = \cos^{-1}\left(1 - \frac{A}{2\pi R_0^2}\right)
\ee
of a spherical cap having the same area $A$ as the plate in question, 
using the plate areas from table 1 of \citet{bird03gcubed}. The
resulting values of $\theta_t$ are given in column 2 of 
table \ref{tab:2}.

To illustrate the subsequent steps, we begin with the Cocos plate subducting
beneath Central America. We used the five trench-normal transects MEX6 and COST1-COST4
from table 1 of \citet{lallemand05gcubed}, which bracket a segment of the
trench about 880 km long. The average values of the slab length and the deep slab dip for these
transects are $l = 550$ km and $\varphi_s = 59^{\circ}$, respectively. The average
age of the lithosphere at the trench is $\tau = 21.4$ Ma. The lithospheric thickness $h$
corresponding to this value of $\tau$ can be obtained from the standard
half-space cooling temperature profile
\be
T = T_0 + \Delta T\mathrm{erf}\frac{z}{2\sqrt{\kappa\tau}}
\label{hscool}
\ee
where $T_0 = 0^{\circ}$C, $\Delta T = 1325^{\circ}$C, $\kappa = 8\times 10^{-7}$ m$^2$ s$^{-1}$ is the thermal diffusivity, and $z$ is the depth from the upper surface of the lithosphere. Defining
$h$ as the depth to the 1200$^{\circ}$C isotherm, we find $h = 2\sqrt{\kappa\tau}\mathrm{erf}^{-1}0.9057 = 2.36 \sqrt{\kappa\tau}=55$ km. The dimensionless slab
length is therefore $l/h = 10.0$, whence our slab geometry model (\ref{slabshape})
implies $\theta_s = 12.2^{\circ}$ given $\theta_t = 8.65^{\circ}$.
The values of $\tau$, $h$, $l$, $\varphi_s$ and $\theta_s$ are given
in columns 4-8 of table \ref{tab:2}.

The calculations for the remaining subduction zones proceed similarly.
In each case, we used 3 to 5 neighbouring transects from 
table 1 of \citet{lallemand05gcubed}. These transects are TONG1-TONG4 for
Tonga; NMAR1-NMAR4 for Marianas; NCHI3-NCHI6 for Chile; 
RYU1-RYU5 for Ryukyu; and CASC2-CASC4 for Cascadia. For Tonga and
Marianas, $h$ cannot be calculated using the halfspace
cooling model Eq.~\eqref{hscool}, which
ceases to be valid at $\tau\approx 70$ Ma \citep{parsons77jgr}. 
For greater ages the thickness of the lithosphere tends to a constant asymptotic value
$h\approx 100$ km, and we therefore used this value for Tonga and Marianas.

\begin{table}
	\centering
	\begin{tabular}{ccccccccc}
		\hline
		Plate & $\theta_t$ (deg.) & Subduction zone & $\tau$ (Ma) &  $h$ (km) & $l$ (km) & $\varphi_s$ (deg.) & $\theta_s$ (deg.) \\
		\hline
		PA  & 53.5 & Tonga & 107 & 100 & 890 & 54 & 59.7 \\
		PA  & 53.5 & Marianas & 148 & 100 & 770 & 82 & 55.8\\
		NZ  & 20.5 & Chile & 53.5 & 86.7 & 1200 & 45 & 29.8 \\
		PS  & 11.9 &  Ryukyu & 44.2 & 77.8 & 590 & 61 & 15.5  \\
		CO  & 8.7  &  Cent. America & 21.4 & 55.0 & 550 & 59 & 12.2 \\
		JF  & 2.6 & Cascadia & 10.7 & 38.7 & 730 & 45 & 8.2\\
	\end{tabular}
	\caption{Calculated parameters for Pacific subduction zones}
	\label{tab:2}
\end{table}


\bibliographystyle{cambridgeauthordate}

\bibliography{ribe}

\end{document}